\documentclass[twocolumn,epjc3]{svjour3}
\journalname{Eur. Phys. J. C}
\def\thetitle{Measurement of \WZ\ Production in Proton-Proton Collisions at $\sqrt{s}=7$~TeV with the ATLAS Detector}
\title{\thetitle}
\titlerunning{\thetitle}
\author{The ATLAS Collaboration}
\institute{CERN, 1211 Geneva 23, Switzerland}

\usepackage{preprintcover}
\PreprintCoverPaperTitle{\thetitle} 
\PreprintIdNumber{CERN-PH-EP-2012-179}  
\PreprintCoverAbstract{A study of \WZ\ production in proton-proton collisions at $\sqrt{s} = 7$~TeV
is presented using data corresponding to an integrated luminosity of \lumiRound\,\ifb\
collected with the ATLAS detector at the Large Hadron Collider in 2011.
In total, \WZNSigEventsObserved\ candidates, with a background expectation of
$\WZNBkgEventsRound\pm\WZNBkgEventsRoundErr$ events, are observed in double-leptonic decay final states with electrons,
muons and missing transverse momentum.
The total cross-section is determined to be 
$\xsectot = \WZtotXsec\WZtotXsecStatErr\text{(stat.)}\WZtotXsecSysErr\text{(syst.)}\WZtotXsecLumiErr\text{(lumi.)}\,\pb$,
consistent with the Standard Model expectation of $\WZtotXsecExpected\WZtotXsecExpectedErr\,\pb$.
Limits on anomalous triple gauge boson couplings are derived using the transverse momentum spectrum of $Z$ bosons in the selected events.
The cross-section is also presented as a function of $Z$ boson transverse momentum and 
diboson invariant mass.}
\PreprintJournalName{European Physical Journal C}  

\usepackage[hyperindex,breaklinks]{hyperref} 
\usepackage{graphicx}
\usepackage{amsmath}
\usepackage{atlasphysics}
\usepackage{epstopdf}
\usepackage{cite}
\usepackage{array}

\newcommand{\xsectot}{\ensuremath{\sigma_{W\!Z}^\mathrm{tot}}}
\newcommand{\xsecfid}{\ensuremath{\sigma_{W\!Z}^\mathrm{fid}}}
\newcommand{\ptZ}{\ensuremath{\pt^Z}}
\newcommand{\mWZ}{\ensuremath{m_{W\!Z}}}

\newcommand{\gZ}{g_1^Z}
\newcommand{\kZ}{\kappa_Z}
\newcommand{\dgZ}{\Delta\gZ}
\newcommand{\dkZ}{\Delta\kZ}
\newcommand{\lamZ}{\lambda_Z}
\newcommand{\Nobs}{N_\mathrm{obs}}
\newcommand{\Nsig}{N_\mathrm{sig}}
\newcommand{\Nbkg}{N_\mathrm{bkg}}
\newcommand{\Zjets}{\ensuremath{Z+\text{jets}}}
\newcommand{\eee}{e^{\pm}e^+e^-}
\newcommand{\mme}{e^{\pm}\mu^+\mu^-}
\newcommand{\eem}{\mu^{\pm}e^+e^-}
\newcommand{\mmm}{\mu^{\pm}\mu^+\mu^-}

\def\lumi{4.64}
\def\lumiAbsErr{0.08}
\def\lumiRelErr{1.8}
\def\lumiRound{4.6}
\def\WZtotXsecExpected{17.6}
\def\WZtotXsecExpectedErr{^{+1.1}_{-1.0}}

\def\WZNSigEventsObserved{317}

\def\WZNSigEventsRound{231}

\def\WZNSigEventsRoundErr{8}

\def\WZNBkgEventsRound{68}

\def\WZNBkgEventsRoundErr{10}
\def\WZNSigEventsObservedEEE{56}
\def\WZNSigEventsExpectedEEE{38.9}

\def\WZNSigEventsExpectedErrEEE{2.1}

\def\WZNBkgEventsExpectedEEE{14.5}

\def\WZNBkgEventsExpectedErrEEE{2.9}

\def\WZNSigEventsObservedEEM{75}
\def\WZNSigEventsExpectedEEM{54.0}

\def\WZNSigEventsExpectedErrEEM{2.2}

\def\WZNBkgEventsExpectedEEM{11.5}

\def\WZNBkgEventsExpectedErrEEM{2.5}

\def\WZNSigEventsObservedEMM{78}
\def\WZNSigEventsExpectedEMM{56.6}

\def\WZNSigEventsExpectedErrEMM{1.7}

\def\WZNBkgEventsExpectedEMM{21.0}

\def\WZNBkgEventsExpectedErrEMM{3.5}

\def\WZNSigEventsObservedMMM{108}
\def\WZNSigEventsExpectedMMM{81.7}

\def\WZNSigEventsExpectedErrMMM{2.1}

\def\WZNBkgEventsExpectedMMM{21.0}

\def\WZNBkgEventsExpectedErrMMM{5.6}

\def\WZtotXsec{19.0}
\def\WZtotXsecStatErr{^{+1.4}_{-1.3}}
\def\WZtotXsecSysErr{\pm0.9}
\def\WZtotXsecLumiErr{\pm0.4}
\def\WZfidXsec{92}
\def\WZfidXsecStatErr{^{+7}_{-6}}
\def\WZfidXsecSysErr{\pm4}
\def\WZfidXsecLumiErr{\pm2}
\def\WZtotXsecMuon{0.7}
\def\WZtotXsecElec{2.0}
\def\WZtotXsecMET{0.5}
\def\WZtotXsecTrig{0.2}
\def\WZtotXsecGenerator{0.4}
\def\WZtotXsecPDF{1.2}
\def\WZtotXsecScale{0.4}
\def\WZtotXsecSigMC{0.5}
\def\WZtotXsecBkgMC{0.4}
\def\WZtotXsecBkgDD{4.0}
\def\WZtotXsecSys{4.8}
\def\WZfidXsecMuon{0.7}
\def\WZfidXsecElec{2.1}
\def\WZfidXsecMET{0.5}
\def\WZfidXsecTrig{0.2}
\def\WZfidXsecSigMC{0.5}
\def\WZfidXsecBkgMC{0.4}
\def\WZfidXsecBkgDD{4.0}
\def\WZfidXsecSys{4.6}
\def\metSysJetResEMM{0.4}

\def\muonSysisoIPMME{0.4}
\def\AWZrelErrorGenerator{0.4}
\def\muonSysisoIPMMM{0.6}

\def\electronSysIDEEE{3.5}
\def\electronSysIDEEM{2.3}
\def\metSysTopoClusterScaleEMM{0.6}
\def\electronSysEnergySmearingEEE{0.1}
\def\electronTriggerSystematicEEE{<0.1}
\def\electronSysEnergySmearingEEM{0.1}
\def\AWZrelErrorScale{0.4}

\def\electronTriggerSystematicEEM{<0.1}
\def\metSysJetScaleEMM{0.1}

\def\AWZmmm{0.338}

\def\electronSysEnergyScaleEEE{0.5}

\def\electronSysEnergyScaleEEM{0.3}

\def\AWZeee{0.330}
\def\AWZeem{0.332}
\def\metSysPileupMMM{0.1}
\def\muonSysSmearingMEE{<0.1}

\def\metSysPileupEEE{0.3}
\def\metSysPileupEEM{0.1}
\def\electronSysIDEMM{1.2}
\def\muonTriggerSystematicMMM{0.3}

\def\electronSysEnergySmearingEMM{<0.1}

\def\electronTriggerSystematicEMM{<0.1}
\def\AWZrelErrorComb{1.2}

\def\electronSysRecoEEE{2.5}

\def\electronSysRecoEEM{1.7}

\def\muonTriggerSystematicEEM{0.1}

\def\electronSysCaloIsoEEE{1.5}
\def\electronSysCaloIsoEEM{1.1}

\def\muonSysEffMEE{0.3}
\def\electronSysEnergyScaleEMM{0.3}

\def\AWZemm{0.333}
\def\metSysJetResMMM{0.2}
\def\muonSysSmearingMME{0.1}

\def\muonSysSmearingMMM{0.1}

\def\metSysPileupEMM{0.3}
\def\metSysJetResEEE{0.3}
\def\metSysTopoClusterScaleMMM{0.2}
\def\metSysJetResEEM{0.3}

\def\muonSysisoIPMEE{0.2}
\def\metSysJetScaleMMM{0.1}
\def\electronSysRecoEMM{0.8}

\def\metSysTopoClusterScaleEEE{0.4}

\def\metSysTopoClusterScaleEEM{0.2}
\def\muonTriggerSystematicEMM{0.1}

\def\metSysJetScaleEEE{0.1}

\def\metSysJetScaleEEM{0.1}

\def\electronSysCaloIsoEMM{0.4}

\def\muonSysEffMME{0.5}

\def\muonSysEffMMM{0.8}

\def\WZNZZBkgEventsExpected{21.0}

\def\WZNTTWZBkgEventsExpected{4.7}
\def\WZNTTWZBkgEventsExpectedErr{0.2}
\def\WZNZgammaBkgEventsExpected{3.7}
\def\WZNZgammaBkgEventsExpectedErr{1.1}
\def\ttbarRatio{2.2}
\def\ttbarRatioErr{1.0}

\def\WZNBkgEventsExpectedEEE{14.5}
\def\WZNBkgEventsExpectedErrEEE{2.9}
\def\WZNBkgEventsExpectedEEM{11.5}
\def\WZNBkgEventsExpectedErrEEM{2.5}
\def\WZNBkgEventsExpectedEMM{21.0}
\def\WZNBkgEventsExpectedErrEMM{3.5}
\def\WZNBkgEventsExpectedMMM{21.0}
\def\WZNBkgEventsExpectedErrMMM{5.6}
\def\WZNZjetsDDBkgEventsExpected{31.9}
\def\WZNZjetsDDBkgEventsExpectedErr{9.2}
\def\WZNZjetsDDBkgEventsExpectedEEE{8.8}
\def\WZNZjetsDDBkgEventsExpectedErrEEE{2.8}
\def\WZNZjetsDDBkgEventsExpectedEEM{3.7}
\def\WZNZjetsDDBkgEventsExpectedErrEEM{2.3}
\def\WZNZjetsDDBkgEventsExpectedEMM{10.2}
\def\WZNZjetsDDBkgEventsExpectedErrEMM{3.3}
\def\WZNZjetsDDBkgEventsExpectedMMM{9.1}
\def\WZNZjetsDDBkgEventsExpectedErrMMM{5.5}
\def\WZNZZBkgEventsExpected{21.0}
\def\WZNZZBkgEventsExpectedErr{0.7}
\def\WZNZZBkgEventsExpectedEEE{3.2}
\def\WZNZZBkgEventsExpectedErrEEE{0.2}
\def\WZNZZBkgEventsExpectedEEM{4.9}
\def\WZNZZBkgEventsExpectedErrEEM{0.2}
\def\WZNZZBkgEventsExpectedEMM{5.0}
\def\WZNZZBkgEventsExpectedErrEMM{0.1}
\def\WZNZZBkgEventsExpectedMMM{7.9}
\def\WZNZZBkgEventsExpectedErrMMM{0.2}
\def\WZNTTbarBkgEventsExpected{6.8}
\def\WZNTTbarBkgEventsExpectedErr{3.2}
\def\WZNTTbarBkgEventsExpectedEEE{0.4}
\def\WZNTTbarBkgEventsExpectedErrEEE{0.4}
\def\WZNTTbarBkgEventsExpectedEEM{1.7}
\def\WZNTTbarBkgEventsExpectedErrEEM{0.9}
\def\WZNTTbarBkgEventsExpectedEMM{2.3}
\def\WZNTTbarBkgEventsExpectedErrEMM{1.1}
\def\WZNTTbarBkgEventsExpectedMMM{2.4}
\def\WZNTTbarBkgEventsExpectedErrMMM{1.2}
\def\WZNTTWZBkgEventsExpected{4.7}
\def\WZNTTWZBkgEventsExpectedErr{0.2}
\def\WZNTTWZBkgEventsExpectedEEE{0.7}
\def\WZNTTWZBkgEventsExpectedErrEEE{0.1}
\def\WZNTTWZBkgEventsExpectedEEM{1.2}
\def\WZNTTWZBkgEventsExpectedErrEEM{0.1}
\def\WZNTTWZBkgEventsExpectedEMM{1.3}
\def\WZNTTWZBkgEventsExpectedErrEMM{0.1}
\def\WZNTTWZBkgEventsExpectedMMM{1.6}
\def\WZNTTWZBkgEventsExpectedErrMMM{0.1}
\def\WZNZgammaBkgEventsExpected{3.7}
\def\WZNZgammaBkgEventsExpectedErr{1.1}
\def\WZNZgammaBkgEventsExpectedEEE{1.4}
\def\WZNZgammaBkgEventsExpectedErrEEE{0.7}

\def\WZNZgammaBkgEventsExpectedEMM{2.3}
\def\WZNZgammaBkgEventsExpectedErrEMM{0.9}

\def\obskappalowtwo{-0.42}

\def\obskappaup{0.57}
\def\expkappalow{-0.33}
\def\obslambdalow{-0.046}
\def\obskappauptwo{0.69}

\def\obslambdauptwo{0.066}
\def\obsguptwo{0.133}

\def\obslambdaup{0.047}
\def\expglow{-0.046}
\def\expgup{0.080}

\def\expkappaup{0.47}
\def\explambdalow{-0.041}
\def\obslambdalowtwo{-0.064}
\def\obsglowtwo{-0.074}
\def\explambdaup{0.040}

\def\obsgup{0.093}
\def\obskappalow{-0.37}
\def\obsglow{-0.057}

\begin{document}

\maketitle

\begin{abstract}
A study of \WZ\ production in proton-proton collisions at $\sqrt{s} = 7$~TeV
is presented using data corresponding to an integrated luminosity of \lumiRound\,\ifb\
collected with the ATLAS detector at the Large Hadron Collider in 2011.
In total, \WZNSigEventsObserved\ candidates, with a background expectation of
$\WZNBkgEventsRound\pm\WZNBkgEventsRoundErr$ events, are observed in double-leptonic decay final states with electrons,
muons and missing transverse momentum.
The total cross-section is determined to be 
$\xsectot = \WZtotXsec\WZtotXsecStatErr\text{(stat.)}\WZtotXsecSysErr\text{(syst.)}\WZtotXsecLumiErr\text{(lumi.)}\,\pb$,
consistent with the Standard Model expectation of $\WZtotXsecExpected\WZtotXsecExpectedErr\,\pb$.
Limits on anomalous triple gauge boson couplings are derived using the transverse momentum spectrum of $Z$ bosons in the selected events.
The cross-section is also presented as a function of $Z$ boson transverse momentum and 
diboson invariant mass.
\end{abstract}

\section{Introduction}\label{sec:Introduction}
The underlying structure of electroweak interactions in the Standard Model (SM) is the non-abelian $SU(2)_{L}\times U(1)_{Y}$ gauge group. This model has been very successful in describing measurements to date. Properties of electroweak gauge bosons such as their masses and couplings to fermions have been precisely measured at LEP, the Tevatron and SLD~\cite{Precision}. However, triple gauge boson couplings (TGCs) predicted by this theory have not yet been determined with a similar precision. 

In the SM, the TGC vertex is completely determined by the electroweak gauge structure and so a precise measurement of this vertex, for example through the analysis of diboson production
at the Large Hadron Collider (LHC), tests the gauge symmetry and probes for possible new phenomena involving gauge bosons.
Anomalous TGCs, deviating from gauge constraints, may enhance  the \WZ\ production cross-section  at high diboson invariant masses.
The cross-section can also be enhanced by the production of 
new particles decaying into \WZ\ pairs, 
such as those predicted in supersymmetric models with an extended Higgs sector
and models with extra vector bosons~\cite{DKS:1999}.

\begin{figure}\centering
  {\raisebox{-3mm}{\includegraphics[width=0.33\columnwidth]{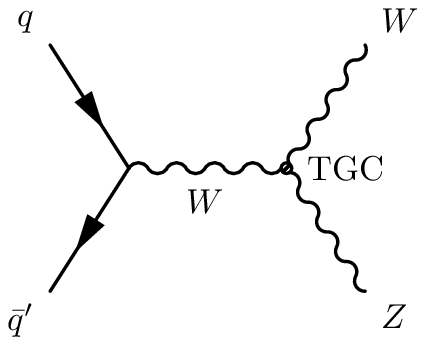}}}\\
  {\includegraphics[width=0.4\columnwidth]{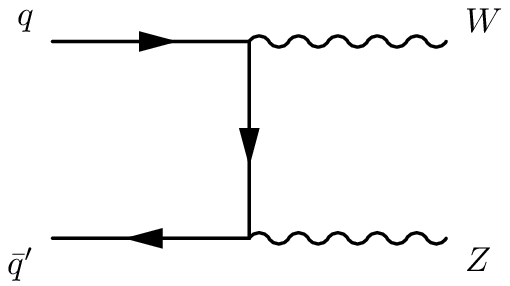}}\quad
  {\includegraphics[width=0.4\columnwidth]{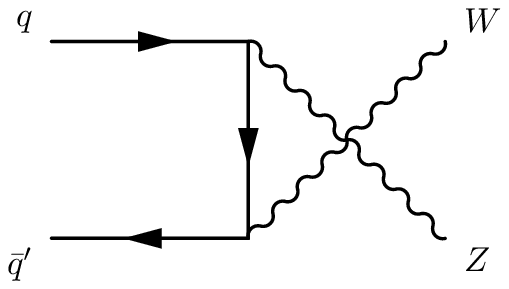}}
\caption{The SM tree-level Feynman diagrams for \WZ production through the $s$-, $t$-, and $u$-channel exchanges in $q\bar{q}'$ interactions at hadron colliders.}
\label{fig:LOdiagrams}
\end{figure}

At the LHC, \WZ\ diboson production arises predominantly from quark-antiquark initial states at leading order (LO) and quark-gluon initial states at next-to-leading order (NLO)~\cite{PhysRevD.65.094041}. 
Figure~\ref{fig:LOdiagrams} shows the LO Feynman diagrams for \WZ production from $q\bar{q}'$ initial states. Only the $s$-channel diagram has a TGC vertex and is hence the only channel to contribute to potential anomalous coupling behaviour of gauge bosons.

In proton-proton ($pp$) collisions at a centre-of-mass energy $\sqrt{s}=7$~TeV,
the SM cross-section for \WZ\ production
is predicted at NLO to be $\WZtotXsecExpected\WZtotXsecExpectedErr$ pb.
This has been computed for $66 < m_{\ell\ell} < 116$~GeV, where $m_{\ell\ell}$ is the
invariant mass of the dilepton system from the $Z$ boson decay, 
using MCFM~\cite{Campbell:2011bn} with the CT10~\cite{ct10}
parton distribution functions (PDFs).
The uncertainty on the prediction comes from the PDF uncertainties, 
evaluated using the CT10 eigenvector sets,
and the QCD renormalization and factorization scales,
which are varied simultaneously up or down by a factor of two
with respect to the nominal value of $(m_W+m_Z)/2$.

This paper presents measurements of the \WZ\ production cross-section
with the ATLAS detector in $pp$ collisions at $\sqrt{s}=7$~TeV. 
The analysis considers four channels of double-leptonic decays \WZl\ 
involving electrons and muons, i.e. $\eee$, $\eem$, $\mme$ and $\mmm$,
plus large missing transverse momentum.
The results are based on an integrated luminosity of $\lumi\pm\lumiAbsErr$\,\ifb\ collected in 2011,
and supersede the earlier ATLAS results based on a subsample of these data~\cite{Aad:2011cx}.

The paper is organized as follows:
Section~\ref{sec:ATLAS} briefly describes the ATLAS detector and the data sample,
including the simulated signal and background samples used in this analysis.
Section~\ref{sec:Reconstruction} details the definition and reconstruction of physically
observable objects such as particles and jets,
and the event selection criteria. 
Section~\ref{sec:SignalAcceptance} presents the signal acceptance, and
Section~\ref{sec:Background} the background estimation.
Section~\ref{sec:Results} presents the measured \WZ\ production cross-section, constraints on the anomalous TGCs,
and the fiducial cross-section as a function of the $Z$ boson transverse momentum and the \WZ\ diboson invariant mass.

\section{The ATLAS Detector and Data Sample}\label{sec:ATLAS}
The ATLAS detector~\cite{atlas} is a multi-purpose particle phys\-ics detector operating at one of
the beam interaction points of the LHC\@.\footnote{ATLAS uses a right-handed coordinate system with its origin at the nominal interaction point (IP) in the centre of the detector and the $z$-axis along the beam pipe. The $x$-axis points from the IP to the centre of the LHC ring, and the $y$-axis points upward. Cylindrical coordinates $(r,\phi)$ are used in the transverse plane, $\phi$ being the azimuthal angle around the beam pipe. The pseudorapidity is defined in terms of the polar angle $\theta$ as $\eta=-\ln\tan(\theta/2)$.}
The innermost part of the detector is a precision tracking system
covering the pseudorapidity range $|\eta| < 2.5$. It consists of silicon pixels,
silicon strips, and straw-tube chambers operating in a 2\,T axial magnetic field supplied by a
superconducting solenoid.
Outside the solenoid are highly segmented electromagnetic and hadronic
calorimeters covering $|\eta| < 4.9$.

The outermost subsystem is a large muon spectrometer covering $|\eta| < 2.7$, which reconstructs muon tracks and
measures their momenta using the azimuthal magnetic field produced by three sets of air-core
superconducting toroids.
This analysis primarily uses the inner detector and the electromagnetic calorimeter to reconstruct
electrons, the inner detector and the muon spectrometer to reconstruct muons, and the
electromagnetic and hadronic calorimeters to reconstruct the missing
momentum transverse to the beam line, $\MET$. The $\MET$ is corrected to account for muons reconstructed
by the inner detector and the muon spectrometer.

$\WZ$ candidate events with multi-lepton final states are selected 
with single-muon or single-electron triggers.
During the 2011 data-taking,
the transverse momentum (\pt) threshold for the single-muon trigger was 18~GeV\@.
The \pt\ threshold for the 
single-electron trigger was initially 20~GeV, and was raised to 22~GeV in the latter 
part of 2011 to cope with increasing instantaneous luminosity. 
For the \WZ events that pass all selection criteria, 
the trigger efficiency is in the range of (96--99)\% depending on the final state being considered.

\subsection{Simulated Event Samples}

Simulated event samples are used to estimate both the signal selection efficiency and some
of the background contributions.
The response of the ATLAS detector is simulated~\cite{ATLASsimu}
using \geant4~\cite{geant4}.

The production of \WZ\ pairs and subsequent decays
are modelled with the \mcatnlo~\cite{Frixione:2002ik,mc@nlo4manual} event generator,
which incorporates NLO QCD matrix elements into the parton 
shower by interfacing to the~\herwig~\cite{Herwig} program.
The CT10~\cite{ct10} PDF set is used.
The underlying event is modelled with the \Jimmy~\cite{Jimmy} program.  

Background processes for \WZ\ signal detection are jets produced in association with \W\ or \Z\ bosons,
\WW\ and \ZZ\ pairs, and top-quark production events.
\alpgen~\cite{alpgen} is used to model the $\W/\Z+\text{jets}$ and Drell-Yan processes for $\W/\Z$ bosons decaying to $e$, $\mu$ and $\tau$ leptons.
Events with multi-jet production from heavy-flavour partons are modelled with \pythiaB~\cite{pythiab}.
The \WW\ and \ZZ\ processes are modelled with \herwig and \pythia~\cite{pythia}, respectively.
The $\W/\Z+\gamma$ and $\ttbar+\W/\Z$ processes are produced with \madgraph~\cite{madgraph}.
The \ttbar\ and single top-quark events are modelled with \mcatnlo.
Whenever LO event generators are used, the cross-sections are corrected 
to NLO or, if available, NNLO matrix element calculations~\cite{Frixione:2002ik,Campbell:1999ah,Anastasiou:2003ds,2002NuPhB,Kauer:2007zz,Campbell:2012dh,Lazopoulos:2008de}. 

\herwig is used to model the hadronization, initial-state radiation and QCD final-state radiation (FSR),  except for the samples generated with \pythia or \madgraph, for which \pythia is used.
\photos~\cite{photos} is used for QED FSR, and \tauola~\cite{tauola} for the $\tau$ lepton decays.

Each simulated sample is divided into subsamples that reflect the changes in the data-taking conditions in 2011.
The average number of interactions per bunch crossing, $\langle\mu\rangle$, increased throughout 2011 with the instantaneous luminosity, and reached a maximum of $17$.
Particles produced in multiple interactions, either coincident with the event of interest
or in neighbouring bunch crossings, are referred to as `pile-up' and are included in the simulation.
The number
of extra interactions in simulated events is adjusted according to the measured $\langle\mu\rangle$ distribution in each data-taking period.

\section{Event Reconstruction and Selection}\label{sec:Reconstruction}
The following event selection criteria are applied to the events
collected with the single-electron or single-muon trigger described in Section~\ref{sec:ATLAS}.
A primary vertex reconstructed from at least three well-reconstructed charged-particle tracks, each with $\pT>400$ MeV, is required in order to remove non-collision background and ensure good object reconstruction. 
If an event contains more than one primary vertex, the vertex with the largest total $\pt^{2}$
of the associated tracks is selected.  

\subsection{Object Reconstruction and Selection}

Events are selected in the \WZl\ channel,
where the $\ell$ are either $e$ or $\mu$.
The physical objects selected are electrons, muons, and  neutrinos that manifest themselves as $\met$.
Contamination from jets, mainly due to semileptonic decays of hadrons or due to misidentification of hadrons as leptons, is suppressed by requiring the electrons and muons to be 
isolated from other reconstructed objects.

Muon candidates are identified by matching tracks reconstructed in the muon spectrometer to tracks reconstructed in the inner detector.
The momentum of the \emph{combined} muon track is calculated from the momenta of the two tracks corrected for the energy loss in the calorimeter.
To identify muons that traverse fewer than two of the three layers of the muon spectrometer, 
inner detector tracks that match at least one track segment in the muon spectrometer
are also included.
Such muons are referred to as \emph{tagged} muons to distinguish them from the combined muons.
The transverse momentum of the muon must be greater than $15$\,GeV and the pseudorapidity $|\eta| < 2.5$, using the full range of the inner detector.
The muon momentum in simulated events is smeared to account for a small difference in resolution 
between data and simulation.
At the closest approach to the primary vertex, the ratio of the transverse impact parameter $d_0$ to its uncertainty (the $d_0$ significance) must be smaller than three,  
and the longitudinal impact parameter $|z_0|$ must be less than 1\,mm.
These requirements reduce contamination from heavy flavour decays.
Isolated muons are selected with a requirement that the scalar sum of the $\pt$ of the tracks within $\Delta R = 0.3$ of the muon, 
where $\Delta R \equiv \sqrt{(\Delta \eta)^2 + (\Delta \phi)^2}$, must be less than 15\% of the muon $\pt$.

Electron candidates are formed by matching clusters found in the electromagnetic calorimeter to tracks reconstructed in the inner detector~\cite{Aad:2011mk}.
The transverse energy $\et$, calculated from the cluster energy and the track direction, must be greater than
15\,\GeV.
The pseudorapidity of the cluster must be in the ranges $|\eta| < 1.37$ or $1.52 < |\eta| < 2.47$ 
to ensure good containment of the electromagnetic shower in the calorimeters.
The lateral and transverse shapes of the cluster must be consistent with those of an electromagnetic shower.
The $d_0$ significance must be smaller than 10, and $|z_0|$ must be less than 1\,mm.
To ensure isolation, the total calorimeter $\et$ in a cone of $\Delta R = 0.3$ around the electron candidate, not including the $\et$ of the candidate itself, must be less than 14\% of the electron $\et$, and the scalar sum of the $\pt$ of the tracks within $\Delta R = 0.3$ of the electron must be less than 13\% of the electron $\pt$.
The calorimeter response is corrected for the additional energy deposited by pile-up.
The electron energy in simulated events is smeared to account for a small difference in resolution
between data and simulation. 
If an electron candidate and a muon candidate are found within $\Delta R = 0.1$ of each other, 
the electron candidate is removed. This mainly removes final-state radiation, where a photon was misidentified as an electron, and also jets from pile-up that were misidentified as electrons. 

The missing transverse momentum \met\ is estimated from reconstructed electrons with $|\eta| < 2.47$, muons with $|\eta |< 2.7$, jets with $|\eta| < 4.9$, as well as clusters of energy in the calorimeter not included in reconstructed objects with $|\eta|<4.5$~\cite{Aad:2011re}.
The clusters are calibrated to the electromagnetic or the hadronic energy scale according to cluster topology.
The expected energy deposit of the muon in the calorimeter is subtracted. 
Jets are reconstructed with the anti-$k_t$ jet-find\-ing algorithm~\cite{Cacciari:2008gp} with radius parameter $R=0.4$, and are calibrated and corrected for detector effects using simulation, which has been tuned and validated with data.
Events that contain jets, with $\pT>20$\,GeV and $|\eta|<4.9$, which are poorly reconstructed as
determined using calorimeter signal timing and shower shape information,
are rejected to improve \met\ resolution.

\subsection{Signal Event Selection}

Events with two leptons of the same flavour and opposite charge with an invariant mass $m_{\ell\ell}$
within 10\,GeV of the $Z$ boson mass are selected. 
This reduces much of the background from multi-jet, top-quark, and \WW\ production.
Figure~\ref{fig:eventsel}(a) shows the $m_{\ell\ell}$ distribution of the $Z$ candidate in events
that pass the complete event selection criteria described in this section, 
except for the $m_{\ell\ell}$ requirement.

\begin{figure}\centering
\raisebox{0.5\columnwidth}{(a)}\includegraphics[width=0.83\columnwidth]{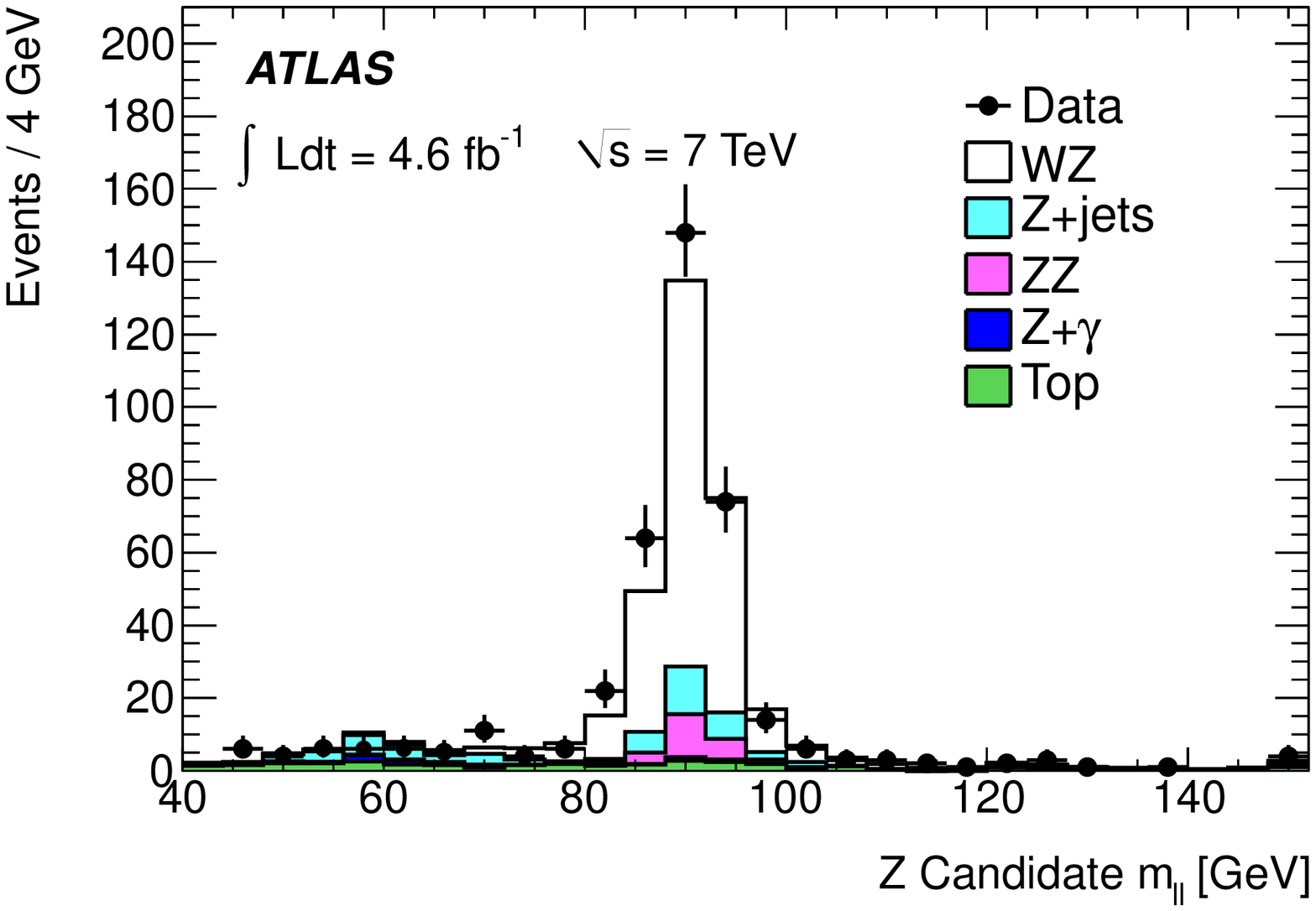}
\raisebox{0.5\columnwidth}{(b)}\includegraphics[width=0.83\columnwidth]{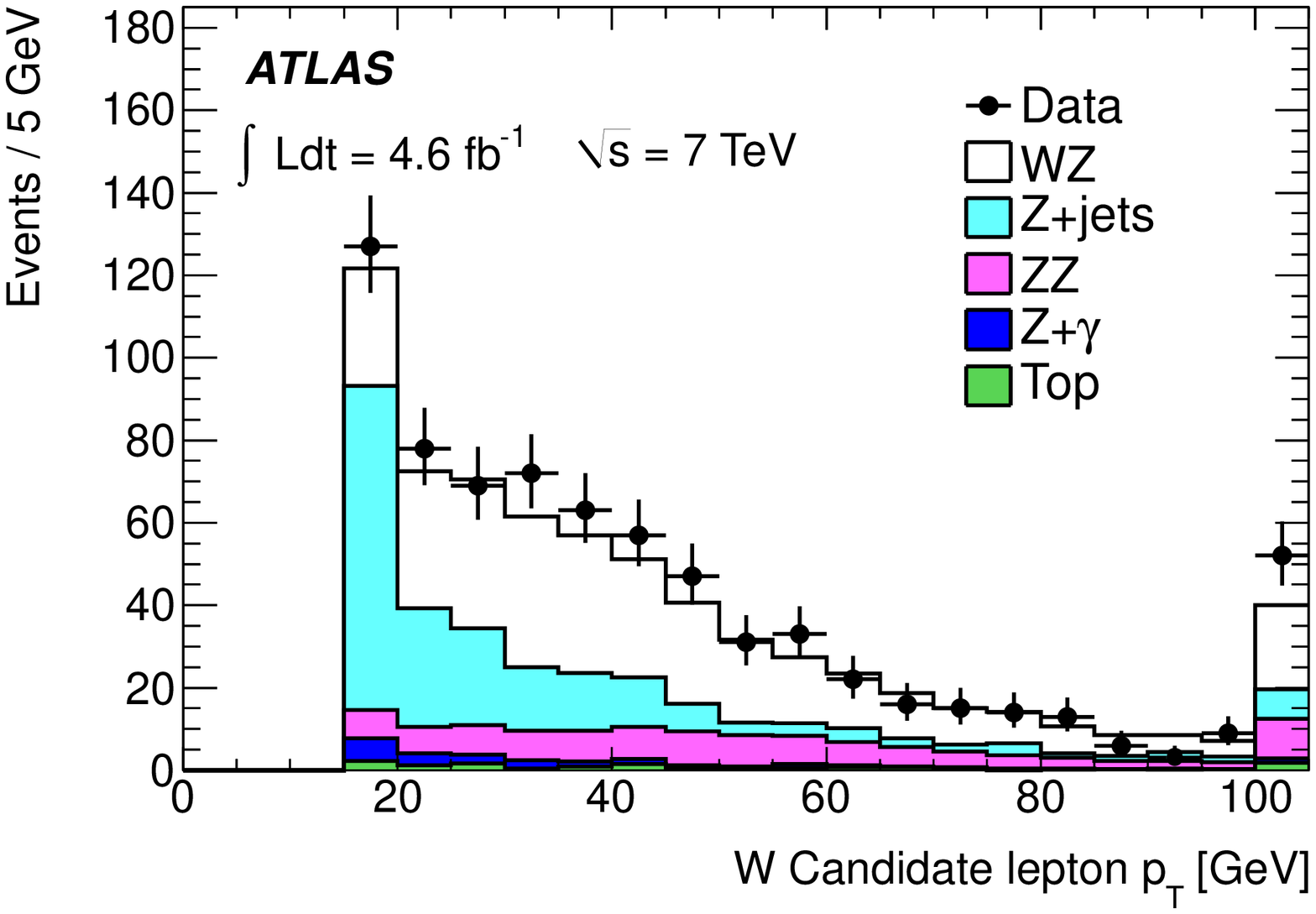}
\raisebox{0.5\columnwidth}{(c)}\includegraphics[width=0.83\columnwidth]{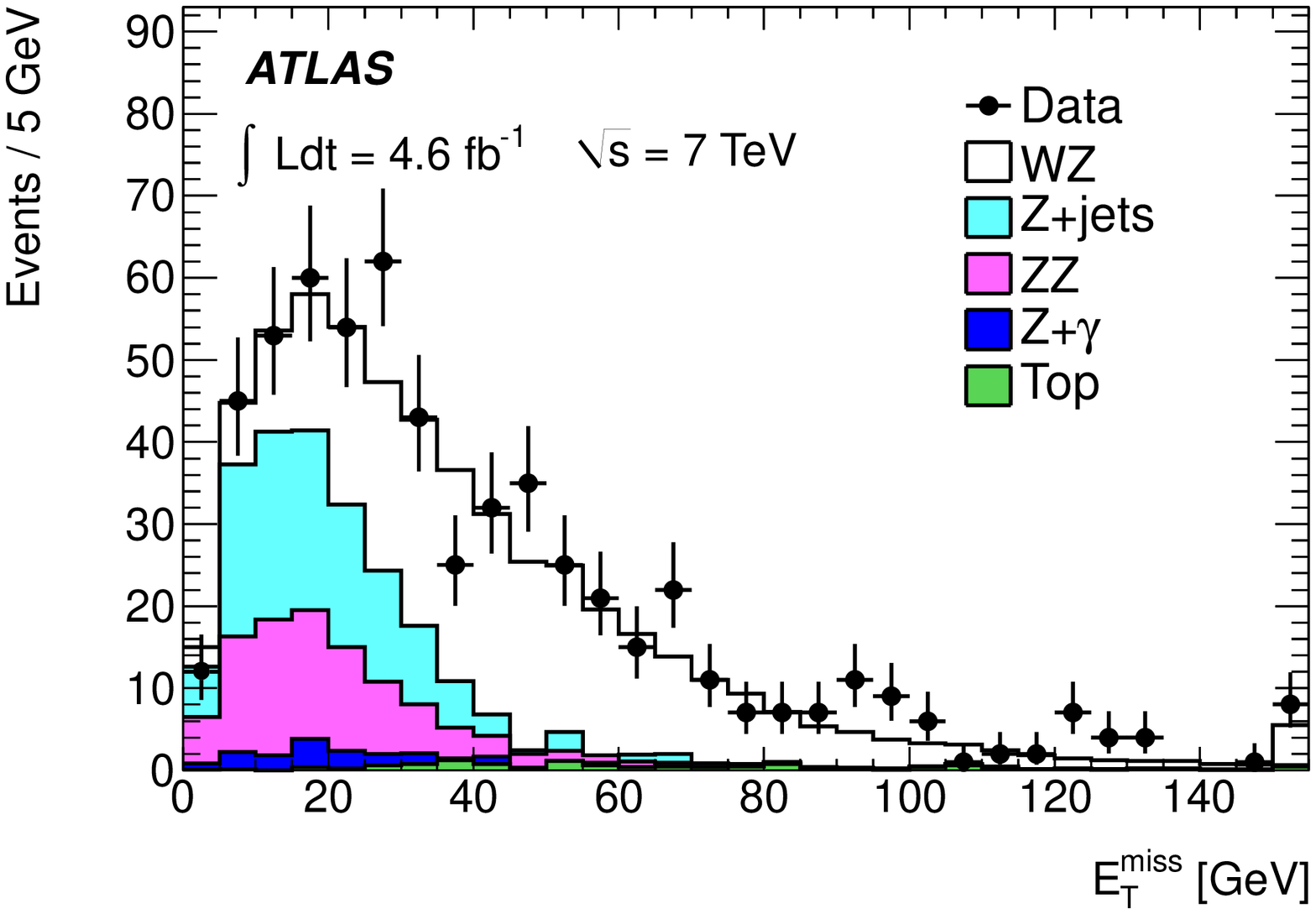}
\raisebox{0.5\columnwidth}{(d)}\includegraphics[width=0.83\columnwidth]{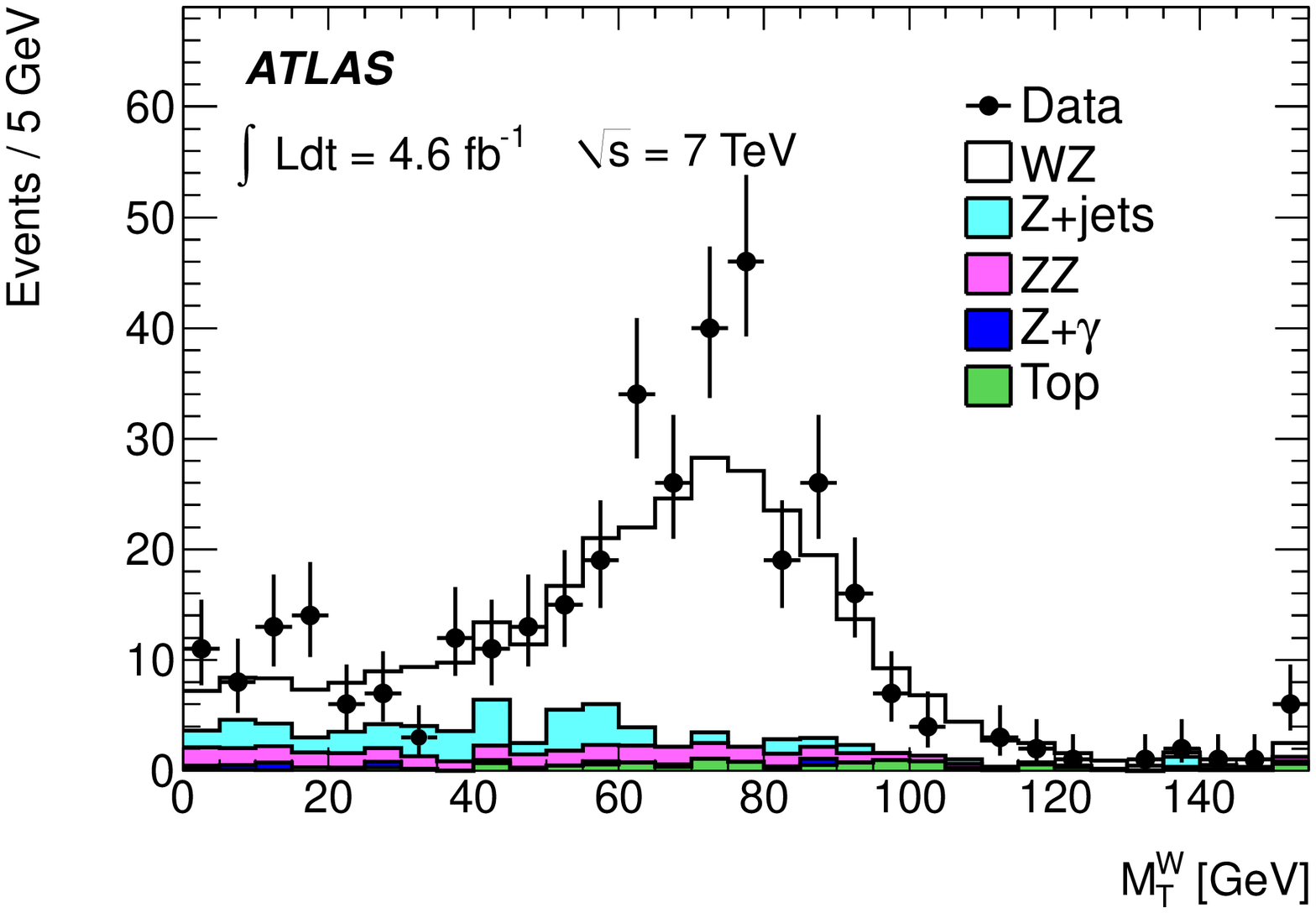}
\caption{%
(a) Dilepton invariant mass $m_{\ell\ell}$ of the $Z$ candidate in the events that pass all event selection criteria
except for the $m_{\ell\ell}$ cut.
(b) Transverse momentum \pt\ of the lepton attributed to the $W$ candidate.
(c) Missing transverse momentum \met\ of the trilepton events.
(d) Transverse mass $\MT^W$ of the $W$ candidates.
Samples shown in (b), (c), and (d) are the candidate events remaining before the cut on
the variable displayed.
The stacked histograms are expectations from simulation for $WZ$, $ZZ$,
and $Z$+$\gamma$. For $Z$+jets
and \ttbar, the expected shape is taken from simulation but the normalization is
taken from the data-driven estimates.  The rightmost bins include overflows.
}\label{fig:eventsel}
\end{figure}

Events are then required to have at least three reconstructed leptons
originating from the same primary vertex,
two leptons from a $Z$ boson decay and one additional  lepton
attributed to the decay of a \W\ boson.
To reduce background from \Zjets,
the third lepton is required to pass more stringent identification criteria 
than required for the leptons attributed to the $Z$ boson.
The additional criteria imposed on electrons are: 
a more stringent quality requirement for the matched track, 
a requirement on the ratio of the energy measured in the calorimeter to the momentum of the matched track, 
and a requirement that transition radiation is detected if the candidate traversed the straw-tube chambers.
Muons attributed to the \W\ boson decay are required to be reconstructed as combined, and not tagged, muons.
Figure~\ref{fig:eventsel}(b) shows the $\pt$ distribution of the third leptons
that pass the additional identification criteria.
The third lepton is required to have $\pt>20$\,GeV.

Figure~\ref{fig:eventsel}(c) shows the $\met$ distribution of the selected trilepton events.
The events have to satisfy $\met$ $>25$\,GeV\@.

The transverse mass of the \W\ boson is calculated as
\begin{equation}
\MT^W = \sqrt{2\pt^\ell\met(1-\cos(\Delta\phi))},
\label{eq:MTW}
\end{equation}
where $\pt^\ell$ is the transverse momentum of the third lepton
and $\Delta\phi$ is the azimuthal angle between the third lepton and the \met.
Figure~\ref{fig:eventsel}(d) shows the $\MT^W$ distribution of the events that reach this stage of the selection.
The observed $\MT^W$ distribution appears to have a narrower peak
than predicted by the simulation.
The events with $70<\MT^W<80$\,GeV have been scrutinized
for signs of experimental problems, and no issues were found.
The limited resolution of the \met\ measurement makes it unlikely that the observed
excess is a narrow peak.

$\MT^W$ is required to be greater than 20\,GeV.
The \met\ and $\MT^W$ requirements  suppress most of
the remaining background from \Zjets\ and other diboson production.

In order to ensure that the trigger efficiency is well determined,
at least one of the muons (electrons) from the $\WZ$
candidate is required to have $\pt>20$ (25) GeV and to be geometrically matched to a muon
(electron) reconstructed by the trigger algorithm.
These \pt\ thresholds are sufficiently large compared with the trigger $\pt$ thresholds to
guarantee that the efficiency is not dependent on the $\pt$ of the lepton.

\section{Signal Acceptance}\label{sec:SignalAcceptance}
The numbers of simulated \WZ\ events after each stage of the event selection, scaled to \lumiRound~\ifb, are listed in Table~\ref{table:cutflowWZMC}.
\begin{table}
	\centering
	\caption{Expected number of signal \WZl events after each
	stage of selection.
	The first nine rows are computed with a simulated \WZ\ sample scaled to \lumiRound\,\ifb.
	Efficiency corrections refer to application of correction factors that account for the
	differences in the trigger and reconstruction efficiencies between the real and simulated data.
	The last row shows the additional contribution from $\WZ\to\tau+X$.}
	\label{table:cutflowWZMC}
	\begin{tabular}{lrrrr}
		\hline
			& $e e e$ & $\mu e e$ &$e \mu \mu $ & $\mu \mu \mu$ \\
		\hline
		Generated & \multicolumn{4}{c}{1202} \\ 
		Muon or electron trigger & \multicolumn{4}{c}{1121} \\ 
		Primary vertex & \multicolumn{4}{c}{1118} \\ 
		Jet cleaning & \multicolumn{4}{c}{1116} \\ 
		Two leptons, $m_{\ell\ell}$   & \multicolumn{2}{c}{219}  & \multicolumn{2}{c}{317} \\
		Three leptons, $\pT$ &  51.2  &  70.6  &  74.8  & 106.6 \\
		$\MET>25$\,GeV &  40.5  &  57.0  &  59.2  &  86.4 \\
		$\MT^W>20$\,GeV &  38.1  &  54.1  &  55.7  &  81.9 \\
		Trigger match &  38.0  &  54.0  &  55.3  &  81.7 \\
		\hline
		Efficiency corrections &  37.2  &  51.8  &  54.2  &  78.3 \\
		\hline
		$\WZ\to\tau X$ contribution & 1.7  &  2.3  &  2.4  &  3.4 \\
		\hline
	\end{tabular}
\end{table}
The ``Efficiency corrections'' row shows the predictions corrected for the differences in the trigger and reconstruction efficiencies between the measured and simulated data.
The acceptance increases with the number of muons in the final state because the reconstruction efficiency for muons is higher than for electrons.
The additional contribution from $\WZ \rightarrow \tau+X$, where the $\tau$ decays into an electron or a muon,
is shown in the last row of Table~\ref{table:cutflowWZMC}.

Table~\ref{table:accSysflavor} summarizes the systematic uncertainties on the expected signal yields.
For electrons and muons,
the reconstruction efficiencies, \pt\ scale and resolution, and
efficiencies for the isolation and impact-parameter requirements
are studied using samples of \W, \Z, and $J/\psi$ decays.
Differences observed between data and simulated samples are accounted for,
and the uncertainties in the correction factors are used to evaluate the
systematic uncertainties.

The uncertainties related to \met\ come mainly from the calibration of 
cluster and jet energies.
The effects of event pile-up are evaluated from the distribution of
total transverse energy as a function of $\langle\mu\rangle$.

Single-muon and single-electron trigger efficiencies are studied in
samples of $Z\to\ell\ell$ events.
Their effects on the \WZ\ measurement are small because the presence
of three leptons provides redundancy for triggering.

The uncertainty in acceptance due to theoretical modelling in the event generator is estimated by comparing
\mcatnlo\ with another NLO generator, \texttt{POWHEG BOX}~\cite{PowhegBox}.
Uncertainties due to the PDFs are computed using the CT10 eigenvectors
and the difference between the CT10 and MSTW 2008~\cite{MSTW2008} PDF sets.
Uncertainties related to the factorization scale $\mu_F$ and renormalization scale $\mu_R$ are estimated by setting $\mu_F=\mu_R$ and varying this value up and down by a factor of two.

\begin{table}
   \centering
   \caption{Systematic uncertainties, in \%, on the expected signal yields.}
   \label{table:accSysflavor}
   \begin{tabular}{l@{}>{$}r<{$}@{ \,}>{$}r<{$}@{ \,}>{$}r<{$}@{\quad}>{$}r<{$}}
   \hline
   Source & eee & \mu ee& e\mu\mu & \mu\mu\mu \\
   \hline
   $\mu$ reconstruction efficiency & - & \muonSysEffMEE & \muonSysEffMME & \muonSysEffMMM \\
   $\mu$ \pT\ scale \& resolution & - & \muonSysSmearingMEE & \muonSysSmearingMME & \muonSysSmearingMMM \\
   $\mu$ isolation \& impact param. & - & \muonSysisoIPMEE & \muonSysisoIPMME & \muonSysisoIPMMM \\
   $e$ reconstruction efficiency & \electronSysRecoEEE & \electronSysRecoEEM & \electronSysRecoEMM & - \\
   $e$ identification efficiency & \electronSysIDEEE & \electronSysIDEEM & \electronSysIDEMM & - \\
   $e$ isolation \& impact param. & \electronSysCaloIsoEEE & \electronSysCaloIsoEEM & \electronSysCaloIsoEMM & - \\
   $e$ energy scale & \electronSysEnergyScaleEEE & \electronSysEnergyScaleEEM & \electronSysEnergyScaleEMM & - \\
   $e$ energy resolution & \electronSysEnergySmearingEEE & \electronSysEnergySmearingEEM & \electronSysEnergySmearingEMM & - \\
   $\MET$ cluster energy scale & \metSysTopoClusterScaleEEE & \metSysTopoClusterScaleEEM &\metSysTopoClusterScaleEMM &\metSysTopoClusterScaleMMM \\
   $\MET$ jet energy scale & \metSysJetScaleEEE & \metSysJetScaleEEM & \metSysJetScaleEMM & \metSysJetScaleMMM \\
   $\MET$ jet energy resolution & \metSysJetResEEE & \metSysJetResEEM & \metSysJetResEMM & \metSysJetResMMM \\
   $\MET$ pile-up & \metSysPileupEEE & \metSysPileupEEM & \metSysPileupEMM & \metSysPileupMMM \\
   Muon trigger & - & \muonTriggerSystematicEEM & \muonTriggerSystematicEMM & \muonTriggerSystematicMMM \\
   Electron trigger & \electronTriggerSystematicEEE & \electronTriggerSystematicEEM & \electronTriggerSystematicEMM & - \\
   \hline
   Event generator & \AWZrelErrorGenerator & \AWZrelErrorGenerator & \AWZrelErrorGenerator & \AWZrelErrorGenerator \\
   PDF & \AWZrelErrorComb & \AWZrelErrorComb & \AWZrelErrorComb & \AWZrelErrorComb \\
   QCD scale & \AWZrelErrorScale & \AWZrelErrorScale & \AWZrelErrorScale & \AWZrelErrorScale\\
   \hline
   Luminosity & \lumiRelErr & \lumiRelErr & \lumiRelErr & \lumiRelErr \\
   \hline
   \end{tabular}
\end{table}

\section{Background Estimation}\label{sec:Background}
The major sources of background are
summarized in Table~\ref{table:BGsummary}.
\begin{table}[b]
  \centering
  \caption{Estimated numbers of background events.
  The errors include both statistical and systematic uncertainties.}
  \label{table:BGsummary}
  \begin{tabular}{l@{\quad}>{$}r<{$}@{\quad}>{$}r<{$}@{\quad}>{$}r<{$}@{\quad}>{$}r<{$}} 
    \hline
    Source & \multicolumn{1}{c}{$eee$} & \multicolumn{1}{c}{$\mu ee$}
           & \multicolumn{1}{c}{$e\mu\mu$} & \multicolumn{1}{c}{$\mu\mu\mu$} \\
    \hline
    $\Zjets$        & \WZNZjetsDDBkgEventsExpectedEEE\pm\WZNZjetsDDBkgEventsExpectedErrEEE
                    & \WZNZjetsDDBkgEventsExpectedEEM\pm\WZNZjetsDDBkgEventsExpectedErrEEM
                    & \WZNZjetsDDBkgEventsExpectedEMM\pm\WZNZjetsDDBkgEventsExpectedErrEMM
                    & \WZNZjetsDDBkgEventsExpectedMMM\pm\WZNZjetsDDBkgEventsExpectedErrMMM \\
    $ZZ$            & \WZNZZBkgEventsExpectedEEE\pm\WZNZZBkgEventsExpectedErrEEE
                    & \WZNZZBkgEventsExpectedEEM\pm\WZNZZBkgEventsExpectedErrEEM
                    & \WZNZZBkgEventsExpectedEMM\pm\WZNZZBkgEventsExpectedErrEMM
                    & \WZNZZBkgEventsExpectedMMM\pm\WZNZZBkgEventsExpectedErrMMM \\
    $Z+\gamma$    & \WZNZgammaBkgEventsExpectedEEE\pm\WZNZgammaBkgEventsExpectedErrEEE
                    & \phantom{0.0}-\phantom{0.0}
                    & \WZNZgammaBkgEventsExpectedEMM\pm\WZNZgammaBkgEventsExpectedErrEMM
                    & \phantom{0.0}-\phantom{0.0} \\
    $\ttbar$        & \WZNTTbarBkgEventsExpectedEEE\pm\WZNTTbarBkgEventsExpectedErrEEE
                    & \WZNTTbarBkgEventsExpectedEEM\pm\WZNTTbarBkgEventsExpectedErrEEM
                    & \WZNTTbarBkgEventsExpectedEMM\pm\WZNTTbarBkgEventsExpectedErrEMM
                    & \WZNTTbarBkgEventsExpectedMMM\pm\WZNTTbarBkgEventsExpectedErrMMM \\
    $\ttbar+W/Z$    & \WZNTTWZBkgEventsExpectedEEE\pm\WZNTTWZBkgEventsExpectedErrEEE
                    & \WZNTTWZBkgEventsExpectedEEM\pm\WZNTTWZBkgEventsExpectedErrEEM
                    & \WZNTTWZBkgEventsExpectedEMM\pm\WZNTTWZBkgEventsExpectedErrEMM
                    & \WZNTTWZBkgEventsExpectedMMM\pm\WZNTTWZBkgEventsExpectedErrMMM \\
    \hline
    Total           & \WZNBkgEventsExpectedEEE\pm\WZNBkgEventsExpectedErrEEE
                    & \WZNBkgEventsExpectedEEM\pm\WZNBkgEventsExpectedErrEEM
                    & \WZNBkgEventsExpectedEMM\pm\WZNBkgEventsExpectedErrEMM
                    & \WZNBkgEventsExpectedMMM\pm\WZNBkgEventsExpectedErrMMM \\
    \hline
  \end{tabular}
\end{table}
Data-driven methods are used to estimate the background from \Zjets\ and $\ttbar$ production.
Simulation is used for the remaining background sources, including $ZZ$, $\ttbar+W/Z$,
and $Z+\gamma$ production.
Background from other sources, such as heavy-flavour multi-jet events, is strongly suppressed by the requirement of three leptons with small $d_0$, and is negligible.
For studies of anomalous TGC (Section~\ref{sec:TGC}) and of normalized fiducial cross-sections (Section~\ref{sec:Unfolding}), the background is estimated separately in bins of the transverse momentum \ptZ\ of the \Z\ boson and
the invariant mass \mWZ\ of the \WZ\ pair.

For background events with three true leptons from vector boson decays,
the simulation models the acceptance and efficiency of the selection
criteria reliably.
The main background in this category is $ZZ$ production, in which both $Z$ bosons decay leptonically.
The background distributions and acceptances are determined directly from simulation for this process,
and the theoretical cross-section is used for normalization.
The total contribution of the $ZZ$ background is 
$\WZNZZBkgEventsExpected\pm\WZNZZBkgEventsExpectedErr$
events, where the error is dominated by the uncertainty on the theoretical
cross-section.

Weak boson radiation associated with $\ttbar$ production can also produce three or more leptons
and thus constitutes a significant background despite its small production cross-section.
The total contribution of the $\ttbar+W/Z$ background is
$\WZNTTWZBkgEventsExpected\pm\WZNTTWZBkgEventsExpectedErr$ events.
$Z+\gamma$ events can pass the selection criteria if the  photon is misidentified as an electron.
The contribution from this background is estimated from simulation to be
$\WZNZgammaBkgEventsExpected\pm\WZNZgammaBkgEventsExpectedErr$
events.

\subsection{$Z+\text{Jets}$ Background}

Production of a \Z\ boson associated with jets is the largest source of background in this measurement.
For a \Zjets\ event to pass the event selection criteria, an isolated lepton must be
reconstructed from one of the jets.
The extra lepton is usually attributed to the \W\ boson.

A \emph{lepton-like jet} is defined as a jet that passes a few basic
lepton selection criteria but not necessarily the full set of selection
 (for $e$) or isolation (for both $e$ and $\mu$) requirements.
An event containing a \Z\ boson and a lepton-like jet is a background event if the lepton-like
jet passes all lepton selection criteria.
Those that fail the lepton quality or isolation requirements constitute a control sample.
To ensure that the control sample is as similar to the signal as possible, all other event selection criteria,  including $\met>25$\,GeV, are applied.

In order to estimate the \Zjets\ background from this control sample,
the probability $f$ of a lepton-like jet passing all lepton selection criteria 
is estimated in another control sample: events containing a $Z$ boson and a lepton-like jet
 with $\MET<25$\,GeV\@. 
This sample is dominated by \Zjets\ events, and  $f$ can thus be directly measured.
The contributions from other processes are subtracted using simulation.
Simulation is also used to estimate the fraction of the \Zjets\ background  in which
a lepton-like jet is attributed to the \Z\ boson.

From the combination of the two control samples, the total \Zjets\ background is estimated to be
$\WZNZjetsDDBkgEventsExpected\pm\WZNZjetsDDBkgEventsExpectedErr$ events.
The largest source of uncertainty is the extrapolation from the $\met<25$\,GeV sample
to the $\met>25$\,GeV sample, which was studied in simulation and in dijet data.
Also included are the statistical uncertainties of the control samples
and the uncertainties on the theoretical cross-sections of the processes subtracted from the control samples.

\subsection{\ttbar\ Background}

A large part of the top-quark background is eliminated by the impact-parameter and isolation requirements
on the leptons, both of which reject lepton candidates originating in jets.
The rejection factors, however, cannot be reliably derived from simulation,
and therefore data-driven corrections are applied to the simulated \ttbar\ events to estimate them.

In this analysis, \ttbar\ events are the only significant source of background that does not
contain a  \Z\ boson.
A control sample enriched in $\ttbar$ background events is defined by changing the charge combination of the
dilepton pair from opposite sign to same sign.
All other selection criteria are unchanged.
Kinematic distributions of simulated $\ttbar$ events are similar in shape and normalization
for same-charge and opposite-charge selections.
The data-to-simulation ratio of the event yield in the same-sign sample is $\ttbarRatio\pm\ttbarRatioErr$.
This ratio is used to scale the $\ttbar$ background predicted in simulation.
Using this procedure, the total contribution from the \ttbar\ background is estimated to be
$\WZNTTbarBkgEventsExpected\pm\WZNTTbarBkgEventsExpectedErr$ events.

\section{Results}\label{sec:Results}
The numbers of expected and observed events after applying all selection criteria are shown in 
Table~\ref{ta:selected_data_MC_chan}.
In total, \WZNSigEventsObserved\, \WZ candidates are observed in data 
with $\WZNSigEventsRound\pm\WZNSigEventsRoundErr$ signal (including final states with $\tau$ leptons)
and $\WZNBkgEventsRound\pm\WZNBkgEventsRoundErr$ background events expected.
There are 206 $W^+Z$ and 111 $W^-Z$ candidates, consistent with the expectations of $186\pm11$ and $110\pm6$, respectively.

\begin{table}
  \centering
  \caption{Summary of observed numbers of events $\Nobs$ and expected signal $\Nsig$ and background $\Nbkg$ contributions. $\Nsig$ includes the contribution from $\WZ\to\tau X$.}
  \label{ta:selected_data_MC_chan}
  \begin{tabular}{l>{$}c<{$}>{$}c<{$}>{$}c<{$}>{$}c<{$}} 
    \hline
                    & eee  & \mu ee & e\mu\mu & \mu\mu\mu \\
   \hline
    $\Nobs$        & \WZNSigEventsObservedEEE 
                    & \WZNSigEventsObservedEEM 
                    & \WZNSigEventsObservedEMM 
                    & \WZNSigEventsObservedMMM \\
    \hline
    $\Nsig$ & \WZNSigEventsExpectedEEE\pm\WZNSigEventsExpectedErrEEE 
                    & \WZNSigEventsExpectedEEM\pm\WZNSigEventsExpectedErrEEM 
                    & \WZNSigEventsExpectedEMM\pm\WZNSigEventsExpectedErrEMM 
                    & \WZNSigEventsExpectedMMM\pm\WZNSigEventsExpectedErrMMM \\
    $\Nbkg$      & \WZNBkgEventsExpectedEEE\pm\WZNBkgEventsExpectedErrEEE
                    & \WZNBkgEventsExpectedEEM\pm\WZNBkgEventsExpectedErrEEM
                    & \WZNBkgEventsExpectedEMM\pm\WZNBkgEventsExpectedErrEMM
                    & \WZNBkgEventsExpectedMMM\pm\WZNBkgEventsExpectedErrMMM \\
    \hline
  \end{tabular}
\end{table}

Figure~\ref{fig:plot-wzcandidate} shows the distributions of the transverse momentum \ptZ\
of the \Z\ boson and the invariant mass \mWZ\ of the \WZ\ pair in selected events.
To reconstruct \mWZ, the momenta of the three leptons are combined with $\met$, and
a \W\ boson mass constraint is used to solve for the $z$ component $p_z^\nu$ of the neutrino momentum.
This generally leaves a two-fold ambiguity which is resolved by choosing
the solution with the smaller $|p_z^\nu|$.
In 27\% of the candidate events, the measured transverse mass is larger than the
nominal \W\ mass, and no real solutions exist for $p_z^\nu$.
The most likely cause is that the measured $\met$ is larger than the actual neutrino \pT. 
In this case, the best estimate is obtained by choosing the real part of the complex solutions,
essentially reducing the magnitude of $\met$ until a physical solution appears.
The data distributions are compared with the expected SM signal and background.
The shapes of the \Zjets\ and \ttbar\  distributions
are taken from simulation and scaled according to the data-driven estimates.
\begin{figure}\centering
\raisebox{0.55\columnwidth}{(a)}\includegraphics[width=0.85\columnwidth]{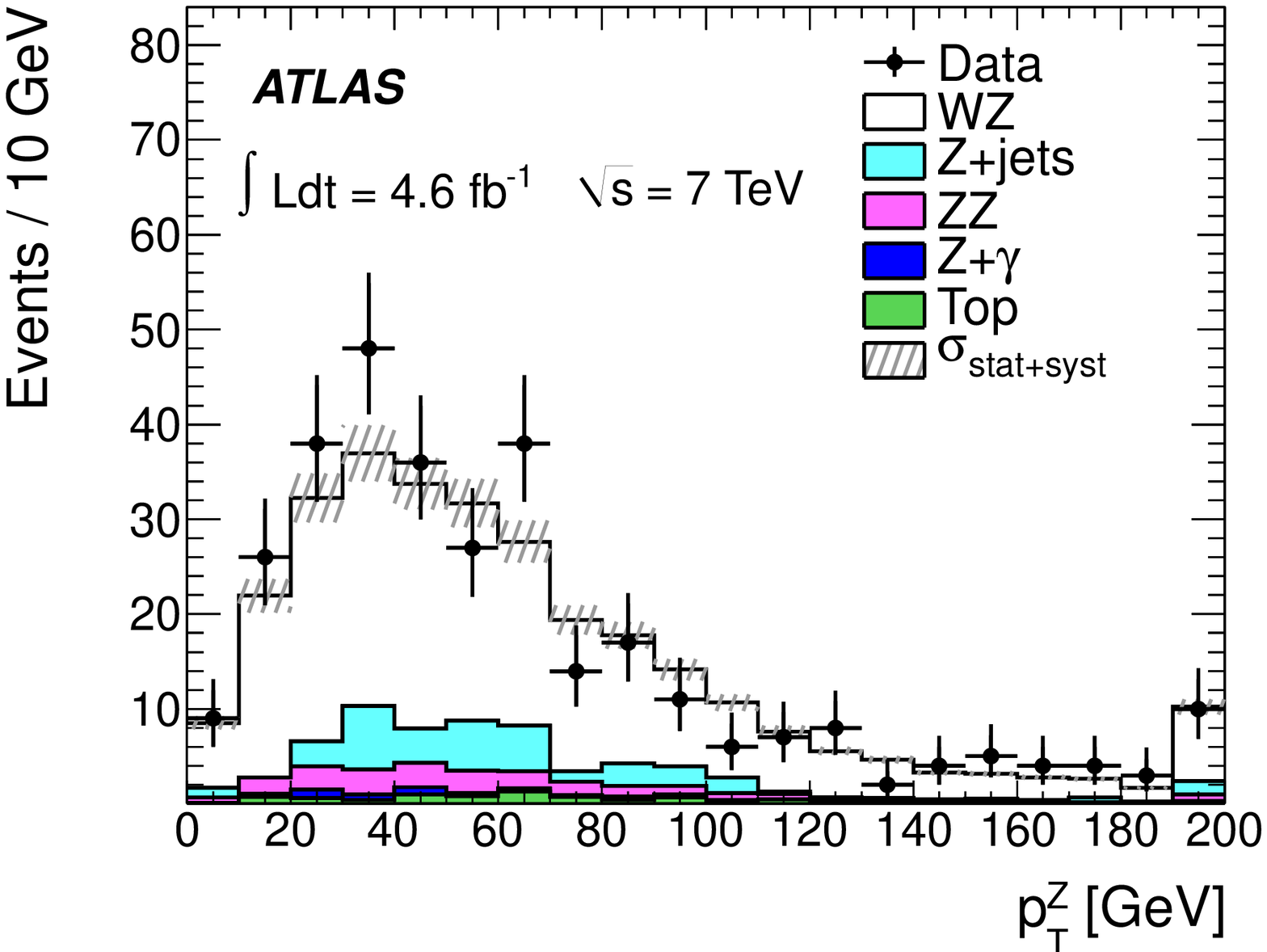}
\raisebox{0.55\columnwidth}{(b)}\includegraphics[width=0.85\columnwidth]{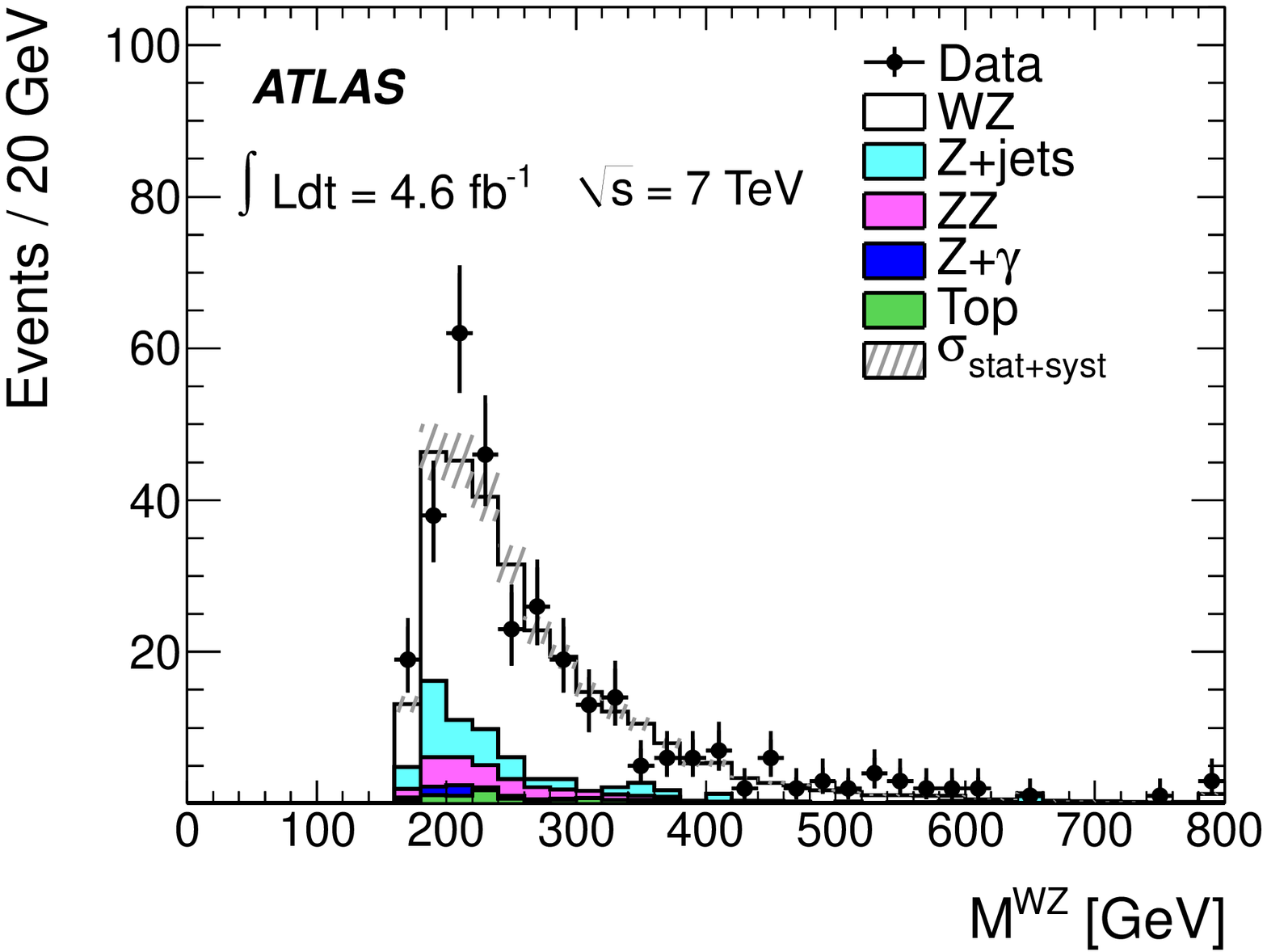}
 \caption{Distributions of the \WZ\ candidates after all selection.
 (a) Transverse momentum $\ptZ$ of the \Z\ boson.
 (b) Invariant  mass  \mWZ\ of the \WZ\ pair.
 The shaded bands indicate the total statistical and systematic
 uncertainties of the prediction. For $Z$+jets 
and \ttbar, the expected shape is taken from simulation but the normalization is
taken from the data-driven estimates.  The rightmost bins include overflow.}
\label{fig:plot-wzcandidate}
\end{figure}

\subsection{Cross-Section Measurement}

Two cross-sections are extracted from the number of observed events.
One is the fiducial \WZl\ cross-section in
a region of final-state phase space defined by the event selection criteria, the other is the total \WZ\ production cross-section.
To extract the total cross-section, theoretical predictions must be used to
extrapolate the measured event yield through the experimentally inaccessible
part of the phase space, introducing additional theoretical
uncertainties. The fiducial cross-section is free from such extrapolation, and is therefore
less sensitive to theoretical uncertainties than the total cross-section.

In order to combine the different channels, a common phase space region is defined in which a fiducial cross-section is extracted. The common phase space is defined as $\pt^{\mu,e} > 15$\,\GeV\ for the leptons from the decay of the \Z\ bosons, $\pt^{\mu,e} > 20$\,\GeV\ for the leptons from the decay of the \W\ bosons, $|\eta^{\mu,e}| < 2.5$, $\pt^{\nu} > 25$\,\GeV, $|m_{\ell\ell}-m_{\rm Z}| < 10$\,\GeV, and 
$M_{\mathrm{T}}^{W} > 20$\,\GeV, to approximate the event selection.
In simulated events, the momenta of photons that are within $\Delta R = 0.1$ of one of the three
leptons are added to the lepton momentum.
In addition, a separation of $\Delta R > 0.3$ between the two leptons of all possible pairings of the three leptons is required.
This requirement emulates the isolation criteria applied to the leptons, which tend to reduce the signal acceptance for events with very large $Z$ boson momenta.
The definition of the fiducial phase space is identical to that used in Ref.~\cite{Aad:2011cx}
except for the requirement of $\Delta R > 0.3$ between the leptons.

\newcommand{\Awz}{A_{W\!Z}}
\newcommand{\Cwz}{C_{W\!Z}}
\newcommand{\NtauMC}{N_\tau^\mathrm{MC}}
\newcommand{\NsigMC}{N_\mathrm{sig}^\mathrm{MC}}
For a given channel \WZl, where $\ell$ is either $e$ or $\mu$, 
the fiducial cross-section is calculated from
\begin{equation}\label{eq:xsecfid}
	\xsecfid
	= \frac{\Nobs - \Nbkg}{\int\mathcal{L}dt \cdot\Cwz}\left(1-\frac{\NtauMC}{\NsigMC}\right),
\end{equation}
where $\Nobs$ and $\Nbkg$ denote the
number of observed and background events respectively, $\int\mathcal{L}dt$
is the integrated luminosity, and $\Cwz$ is the ratio of the number of accepted
signal events to the number of generated events in the fiducial phase space.
Corrections are applied to $\Cwz$ to account for measured differences
in trigger and reconstruction efficiencies between simulated and data samples and for the
extrapolation to the fiducial phase space. 
The contribution from $\tau$ lepton decays, approximately $4\%$, is removed from
the fiducial cross-section definition by the term in  parentheses, where
$\NtauMC$ is the number of accepted simulated \WZ\ events in which at least one of the bosons decays into $\tau$,
and $\NsigMC$ is the number of accepted simulated \WZ\ events with decays into any lepton.
Since the fiducial phase space is defined by the kinematics of the final-state leptons,
the calculated cross-section implicitly includes the leptonic branching fractions of the \W\ and \Z\ bosons. 

\newcommand{\BWlnu}{\mathcal{B}_{W}}
\newcommand{\BZll}{\mathcal{B}_{Z}}
The total cross-section is defined in the dilepton invariant mass range of
$66<m_{\ell\ell}<116$\,GeV for $\Z\to\ell\ell$.
It is computed as: 
\begin{equation}\label{eq:xsectot}
	\xsectot
	= \frac{\Nobs - \Nbkg}{\int\mathcal{L}dt \cdot \BWlnu\BZll
	\Awz\Cwz}\left(1-\frac{\NtauMC}{\NsigMC}\right)
\end{equation}
where $\BWlnu$ and $\BZll$ are the $W$ and $Z$ leptonic branching
fractions, respectively, and
$\Awz$ is the ratio of the number of events within the fiducial phase space to the number of
 events within $66<m_{\ell\ell}<116$\,GeV\@.
The ratio $\Awz$ calculated using \mcatnlo\ equals \AWZeee, \AWZeem, \AWZemm, and \AWZmmm\ for 
the $eee$, $\mu ee$, $e\mu\mu$, and $\mu\mu\mu$ channels, respectively.
The differences are due to the FSR photons emitted outside the $\Delta R=0.1$ cone
around the electrons.

Cross-section measurements are extracted using a maximum-like\-lihood method to combine the four channels.
The likelihood function is defined as
\begin{equation}\label{eq:loglike}
L(\sigma, \vec{x}) 
= \prod_{i=1}^{4} \mathrm{Pois}(\Nobs^i, N_{s}^i(\sigma,\vec{x}) + N_{b}^i(\vec{x}))
 \cdot e^{-\frac{\vec{x}\cdot\vec{x}}{2}}
\end{equation}
where $\mathrm{Pois}(\Nobs^i,N_s^i+N_b^i)$ is the Poisson probability 
of observing $\Nobs^i$ events  in channel $i$
when $N_s^i$ signal and $N_b^i$ background events are  expected.
The nuisance parameters $\vec{x}$ affect $N_s^i$ and $N_b^i$ as
\begin{eqnarray}
N_s^i(\sigma,\vec{x}) &=& N_s^i(\sigma,0) (1 + \sum_k x_k S_k^i), \\
N_b^i(\vec{x}) &=& N_b^i(0) (1 + \sum_k x_k B_k^i),
\end{eqnarray}
where $S_k^i$ and $B_k^i$ are the relative systematic uncertainties on the
signal and background, respectively, due to the $k$-th source of systematic uncertainty.

To find the most probable value of $\sigma$ (fiducial or total) the negative log-likelihood function (from Equation~\ref{eq:loglike}) is minimized simultaneously over $\sigma$ and all the nuisance parameters  $x_k$.
The final results for the combined fiducial and total cross-sections are
\begin{eqnarray*}
\xsecfid &=& \WZfidXsec\WZfidXsecStatErr\text{(stat.)}\WZfidXsecSysErr\text{(syst.)}\WZfidXsecLumiErr\text{(lumi.)}~\text{fb},\\
\xsectot &=& \WZtotXsec\WZtotXsecStatErr\text{(stat.)}\WZtotXsecSysErr\text{(syst.)}\WZtotXsecLumiErr\text{(lumi.)}~\pb.
\end{eqnarray*}
The fiducial cross-section $\xsecfid$ is the sum of the four channels.
Cross-sections extracted separately for the four channels agree within their uncertainties.
The uncertainties are estimated by taking the difference between the
cross-section at the minimum of the negative log-like\-li\-hood
function and the cross-section where the negative log-likelihood is
0.5 units above the minimum in the direction of the fit parameter $\sigma$. The likelihood is maximized over the nuisance parameters for each $\sigma$. 
The systematic uncertainties include all sources except luminosity. 
Correlations between the signal and background uncertainties owing to common sources of systematics
are taken into account in the
definition of the likelihood.
Table~\ref{table:xsecsys} summarizes the systematic uncertainties on the cross-sections from different sources. 
The largest single source of systematic uncertainty is the data-driven estimate of the
background contributions, dominated by that for \Zjets\ production ($\pm3.8\%$). 

\begin{table}
   \centering
   \caption{Systematic uncertainties, in \%, on the fiducial and total cross-sections.
   The background uncertainties are split into data-driven estimates (\Zjets\ and \ttbar)
   and estimates from simulation (all other processes).}
   \label{table:xsecsys}
   \begin{tabular}{l@{}>{$}r<{$}@{ \,}>{$}r<{$}}
   \hline
   Source & \xsecfid & \xsectot \\
   \hline
   $\mu$ reconstruction    & \WZfidXsecMuon & \WZtotXsecMuon \\
   $e$ reconstruction      & \WZfidXsecElec & \WZtotXsecElec \\
   $\MET$ reconstruction   & \WZfidXsecMET & \WZtotXsecMET \\
   Trigger                 & \WZfidXsecTrig & \WZtotXsecTrig \\
   Signal MC statistics    & \WZfidXsecSigMC & \WZtotXsecSigMC \\
   Background data-driven  & \WZfidXsecBkgDD & \WZtotXsecBkgDD \\
   Background MC estimates & \WZfidXsecBkgMC & \WZtotXsecBkgMC \\
   \hline
   Event generator         & - & \WZtotXsecGenerator \\
   PDF                     & - & \WZtotXsecPDF \\
   QCD scale               & - & \WZtotXsecScale \\
   \hline
   Total                   & \WZfidXsecSys & \WZtotXsecSys \\
   \hline
   Luminosity & \lumiRelErr & \lumiRelErr \\
   \hline
   \end{tabular}
\end{table}

\subsection{Anomalous Triple Gauge Couplings}\label{sec:TGC}

General expressions for the effective Lagrangian for the $WWZ$ vertex can be found in Refs.~\cite{Hagiwara,EllisonWudka}.
Retaining only terms that separately conserve charge conjugation $C$ and parity $P$,
the Lagrangian reduces to
\begin{eqnarray}
\label{eqn:Lagrangian}
\frac{\mathcal{L}_{WWZ}}{g_{WWZ}} 
=& i\bigg[ &
\gZ(W^{\dagger}_{\mu\nu}W^{\mu}Z^{\nu}-W_{\mu\nu}W^{\dagger\mu}Z^{\nu}) \nonumber\\
&+& 
\kZ W^{\dagger}_{\mu}W_{\nu}Z^{\mu\nu} +
\frac{\lamZ}{m^{2}_{W}}W^{\dagger}_{\rho\mu}W^{\mu}_{\nu}Z^{\nu\rho}
\bigg]
\end{eqnarray}
where $g_{WWZ}=-e\cot\theta_{W}$, $e$ is the elementary charge,
$\theta_{W}$ is the weak mixing angle,
$W_\mu$ and $Z_\mu$ are the $W$ and $Z$ boson wave functions,
$X_{\mu\nu}\equiv\partial_{\mu}X_{\nu}-\partial_{\nu}X_{\mu}$ for $X=W$ or $Z$,
and $\gZ$, $\kZ$, and $\lamZ$ are dimensionless coupling constants.
The SM predicts $\gZ=1$, $\kZ=1$, and $\lamZ=0$.
This analysis sets limits on possible deviations of these parameters from  their SM values,
i.e.\ on $\dgZ\equiv g_1^Z-1$, $\dkZ\equiv\kappa_Z-1$, and $\lamZ$, known as the anomalous TGC parameters.
The \WZ\ production cross-section is a bilinear function of these anomalous TGCs.

To avoid tree-level unitarity violation,
the anomalous couplings must vanish as $\hat{s}$, the four-momentum squared 
of the \WZ\ system, approaches infinity.
To achieve this, an arbitrary form factor may be introduced~\cite{EllisonWudka}.
Here the dipole form factor adopted is
\begin{equation}
\alpha(\hat{s})=\frac{\alpha_{0}}{(1+\hat{s}/\Lambda^{2})^{2}}
\end{equation}
where $\alpha$ stands for $\dgZ$, $\dkZ$, or $\lamZ$, 
$\alpha_{0}$ is the value of the anomalous coupling at low energy,
and $\Lambda$ is the cut-off scale, the scale at which new physics enters.
The results are reported both with and without this form factor.

Since an enhancement in the cross-section due to an anomalous coupling would grow with $\hat{s}$,
measurement sensitivity to anomalous TGCs is enhanced by binning the data in a kinematic
variable related to $\hat{s}$.
The transverse momentum $\ptZ$ of the $Z$ boson provides a natural choice for
such binning as it is strongly correlated with $\hat{s}$ and can be directly 
reconstructed from the measured lepton momenta with good precision.
The data are therefore divided into six bins in \ptZ\ of width 30\,GeV
followed by a wide bin that includes 180--2000\,GeV.

MC@NLO~\cite{mc@nlo4manual} is used to generate \WZ events with non-SM TGC.\@
The generator computes, for each event, a set of weights that can be used to reweight 
the full sample to any chosen set of anomalous couplings.
This functionality is used to express the predicted signal yields in each bin of $\ptZ$
as a function of the anomalous couplings.
\begin{figure}
\begin{center}
\includegraphics[width=.9\columnwidth]{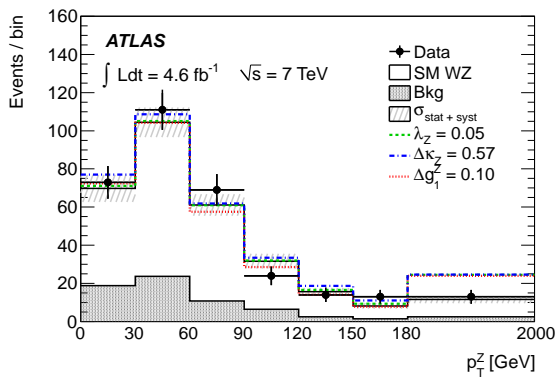}
\caption{Transverse momentum $\ptZ$ of the \Z\ boson in \WZ\ candidate events. 
   Data are shown together with expected background and signal events, 
   assuming the Standard Model.
   Expected events in the case of anomalous TGC without form factor are
   also shown for illustration. 
   The last bin is shortened for display purposes.
}
\label{fig:tgc_ZpT}
\end{center}
\end{figure} 
Figure~\ref{fig:tgc_ZpT} shows the $\ptZ$ distribution of the selected events
together with the SM prediction.
Also shown for illustration are predictions with non-zero anomalous couplings
without form factor:
each coupling is increased to the expected 99\% confidence-level upper limit
while keeping the other two couplings at the SM value.
For this plot the 99\%, rather than 95\%, confidence-level upper limits
are used to accentuate differences in shape.
As expected, the largest deviations from the SM are in the last bin of $\ptZ$,
while the deviations in the lower-\ptZ\ bins depend on which coupling is varied.

Frequentist confidence intervals are obtained on the anomalous couplings
by forming a profile likelihood test incorporating the observed number of candidate events in each $\ptZ$ bin,
the expected signal as a function of the anomalous couplings 
and the estimated number of background events~\cite{frequentist}.
The systematic uncertainties are included in the likelihood function as nuisance parameters
with correlated Gaussian constraints.
A point in the anomalous TGC space is accepted (rejected) at the 95\% confidence level
if less (more) than 95\% of randomly generated pseudo-experiments exhibit
a value of the profile likelihood ratio larger than that observed in data.

\begin{table}\centering
\caption{Expected and observed 95\% confidence intervals on the anomalous couplings
$\dgZ$, $\dkZ$, and $\lamZ$. The expected intervals assume the Standard Model values for the couplings.}
\label{table:limObs}
\begin{tabular}{cccc}
\hline
& Observed & Observed & Expected \\
& $\Lambda=2$~TeV & no form factor & no form factor\\
\hline
$\dgZ$ & $[\obsglowtwo, \obsguptwo]$ & $[\obsglow, \obsgup]$ & $[\expglow, \expgup]$ \\
$\dkZ$ & $[\obskappalowtwo, \obskappauptwo]$ & $[\obskappalow, \obskappaup]$ & $[\expkappalow, \expkappaup]$ \\
$\lamZ$ & $[\obslambdalowtwo, \obslambdauptwo]$ & $[\obslambdalow, \obslambdaup]$ & $[\explambdalow, \explambdaup]$ \\
\hline
\end{tabular}
\end{table}
Table~\ref{table:limObs} summarizes the observed 95\% confidence intervals on the
anomalous couplings $\dgZ$, $\dkZ$, and $\lamZ$, 
with the cut-off scale $\Lambda = 2$~TeV and without
the form factor.
The limits on each anomalous TGC parameter are obtained with the other two anomalous TGC
parameters set to zero.
The expected intervals in Table~\ref{table:limObs} are medians of the
95\% confidence-level upper and lower limits obtained in pseudo-experiments that assume the SM coupling. The widths of the expected and
observed confidence intervals are dominated by statistical uncertainty. 
Figure~\ref{fig:combined} compares the observed limits with the Tevatron results~\cite{CDFlimits,D0limits}.
\begin{figure}\centering
	\includegraphics[width=.9\columnwidth]{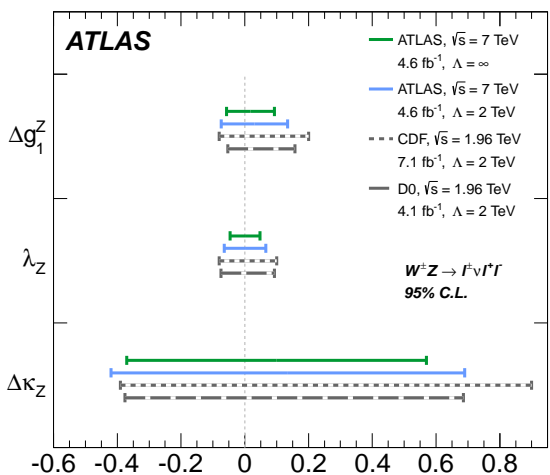}
	\caption{95\% confidence intervals for anomalous TGCs
	from ATLAS (this work), CDF~\cite{CDFlimits}, and D0~\cite{D0limits}.
	Integrated luminosity, centre-of-mass energy and cut-off $\Lambda$ for each experiment are shown.}
	\label{fig:combined}
\end{figure} 

\begin{figure*}\centering
 \includegraphics[width=0.6\columnwidth]{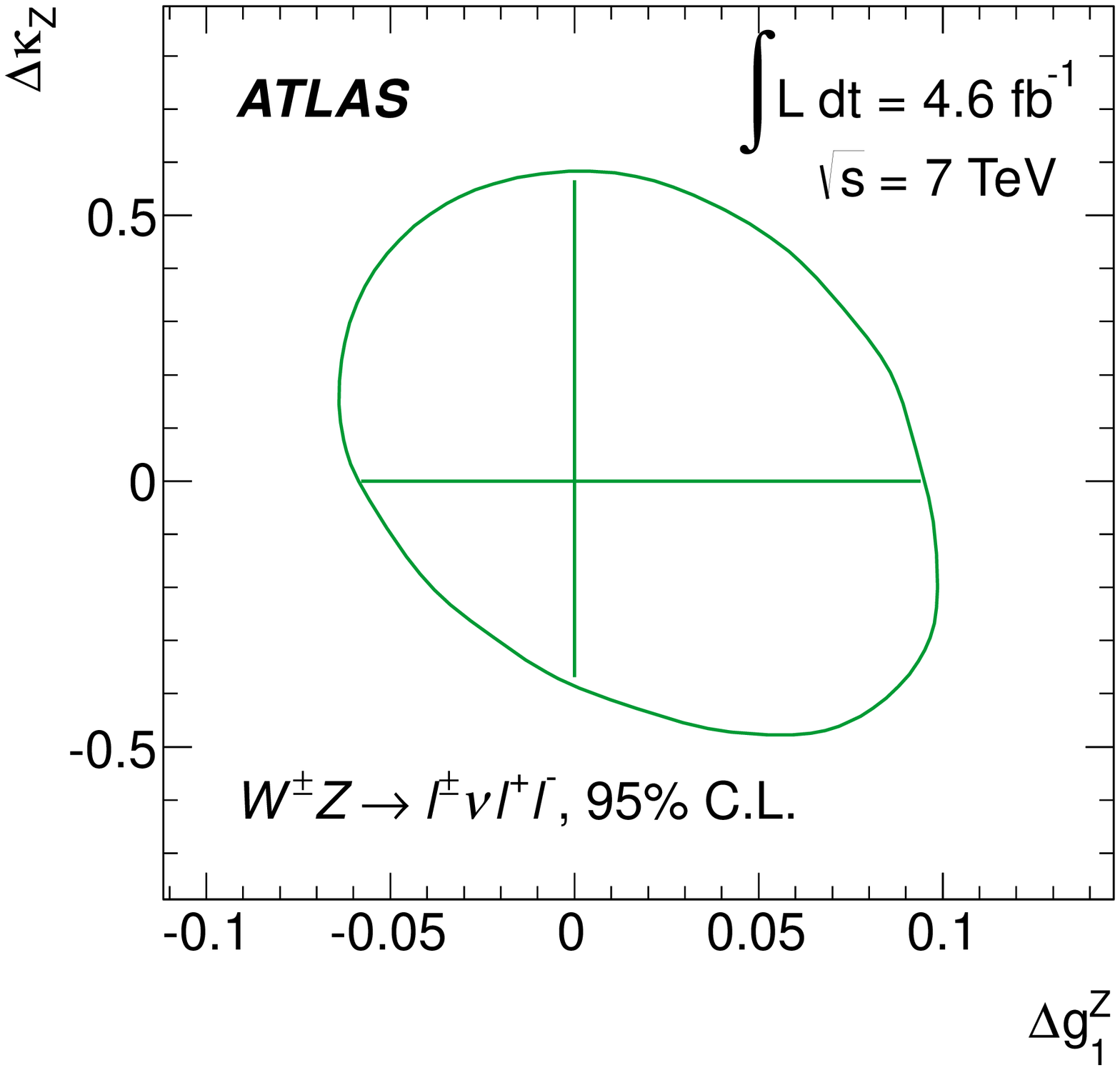}
 \includegraphics[width=0.6\columnwidth]{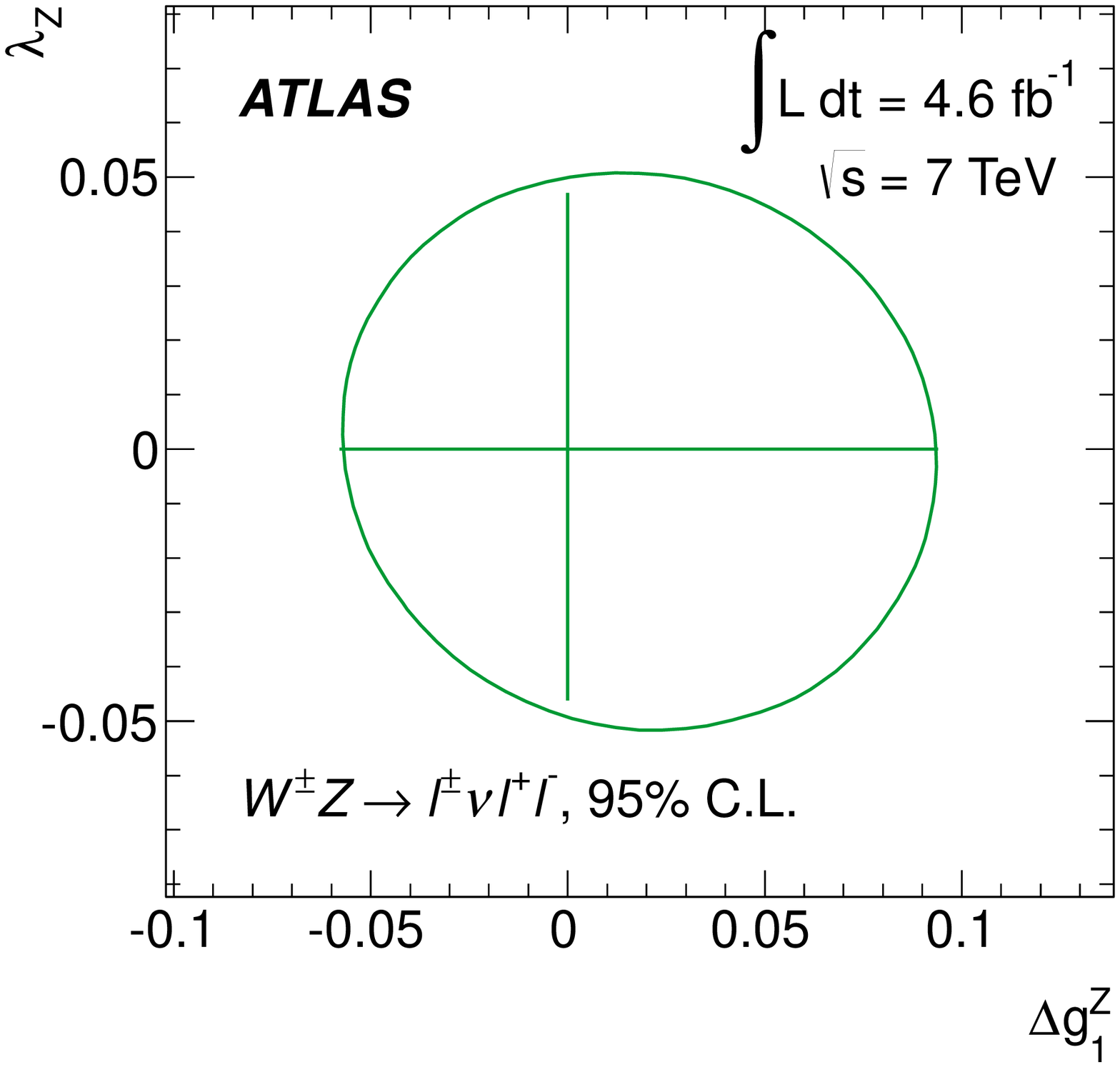}
 \includegraphics[width=0.6\columnwidth]{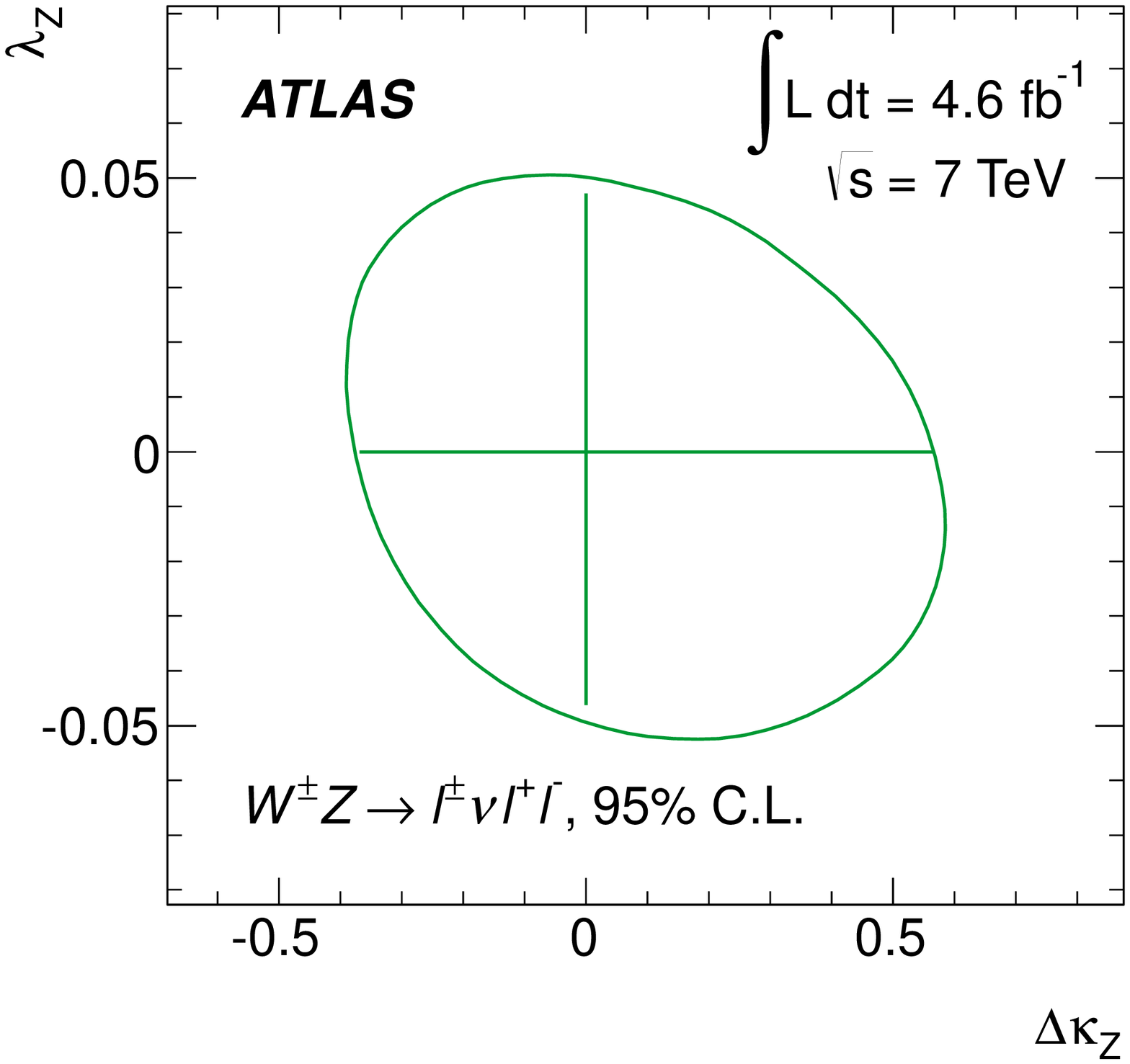}
 \caption{Observed two-dimensional 95\% confidence regions on the
 anomalous couplings without form factor.
 The horizontal and vertical lines inside each
 contour correspond to the limits found in the one-dimensional fit procedure.}
 \label{fig:tgc_2D}
\end{figure*}  
The 95\% confidence regions are shown as contours on the $(\dgZ,\dkZ)$, $(\dgZ,\lamZ)$,
and $(\dkZ,\lamZ)$ planes in Figure~\ref{fig:tgc_2D}.
In each plot the remaining parameter is set to the SM value.
The limits were derived with no form factor.

\subsection{Normalized Fiducial Cross-Sections}\label{sec:Unfolding}

The effective Lagrangian adopted in the TGC analysis in Section~\ref{sec:TGC} allows us
to probe non-SM physics with little model dependence.
An alternative approach is to measure kinematic distributions, such as the \ptZ\ spectrum,
that could be compared with model-dependent theoretical predictions.
For this purpose, it is necessary to convert the measured distributions to the underlying true
distributions by unfolding the effects of the experimental acceptance and resolution.
The iterative Bayesian unfolding proposed by D'Agostini~\cite{iterDAgostini} is applied here. 
An implementation of this technique has previously been used by ATLAS to unfold the 
\pt\ spectrum of inclusively produced \W\ bosons~\cite{Aad:2011fp}.

In the unfolding of binned data, effects of the experimental acceptance and resolution are
expressed in a response matrix, each element of which is the probability of an event in the $i$-th true bin
being reconstructed in the $j$-th measured bin.
In iterative Bayesian unfolding, the response matrix is combined with the measured spectrum
to form a likelihood, which is then multiplied by a prior distribution to produce
the posterior probability of the true spectrum.
The SM prediction is used as the initial prior, and once the posterior probability is obtained, it is
used as the prior for the next iteration after smoothing.
The spectrum becomes insensitive to the initial prior after a few iterations.
The number of iterations is adjusted to control the degree of regularization~\cite{iterDAgostini}.
The differences between successive iterations can be used to estimate the
stability of the unfolding method.

To achieve stable unfolding, that is, without excessive sensitivity
to statistical fluctuations in data or to details of the unfolding technique,
the measured quantity must be a good approximation to the underlying true quantity:
the response matrix must be close to  diagonal.
The \ptZ\ distribution used in the TGC analysis is a natural choice that has
good resolution and sensitivity to possible new physics.
The fractions of \WZ\ events that migrate between two \ptZ\ bins are 2--7\%.

In addition to \ptZ, the distribution of the diboson invariant mass \mWZ\ is also measured.
The resolution of the reconstructed \mWZ\ is limited by the \met\ resolution.
To avoid large bin-to-bin migration and achieve stable unfolding,
three \mWZ\ bins are used: 170--270\,GeV, 270--405\,GeV, and 405--2500\,GeV.
With this binning, the fractions of events that migrate between two \mWZ\ bins are 13--17\%.

Figure~\ref{Fig:Unfold:Data} shows the fiducial cross-sections extracted
in bins of \ptZ\ and \mWZ, normalized by the sum of all bins.
Comparison with the SM prediction shows good agreement.
The corresponding numerical values are presented
in Tables~\ref{Tag:Unfold:Results1PtZ} and~\ref{Tag:Unfold:Results1mWZ} for \ptZ\ and \mWZ, respectively. 

\begin{table*}\centering
\caption{Normalized fiducial cross-sections and uncertainties in bins of \ptZ.}
\label{Tag:Unfold:Results1PtZ} 
\begin{tabular}{l>{$}c<{$}>{$}c<{$}>{$}c<{$}>{$}c<{$}>{$}c<{$}>{$}c<{$}>{$}c<{$}}
\hline
 \ptZ\ [GeV] & [0,30] & [30,60] & [60,90] & [90,120] & [120,150] & [150,180] & [180,2000] \\
\hline
 $\Delta\xsecfid(\ptZ)/\xsecfid$ & 0.231 & 0.350 & 0.230 & 0.065 & 0.045 & 0.042 & 0.038 \\
 Uncertainty  & 0.034 & 0.039 & 0.033 & 0.019 & 0.015 & 0.014 & 0.013 \\
\hline
\end{tabular}
\end{table*}

\begin{figure}
    \begin{center}
    	\raisebox{0.6\columnwidth}{(a)}\includegraphics[width=0.9\columnwidth]{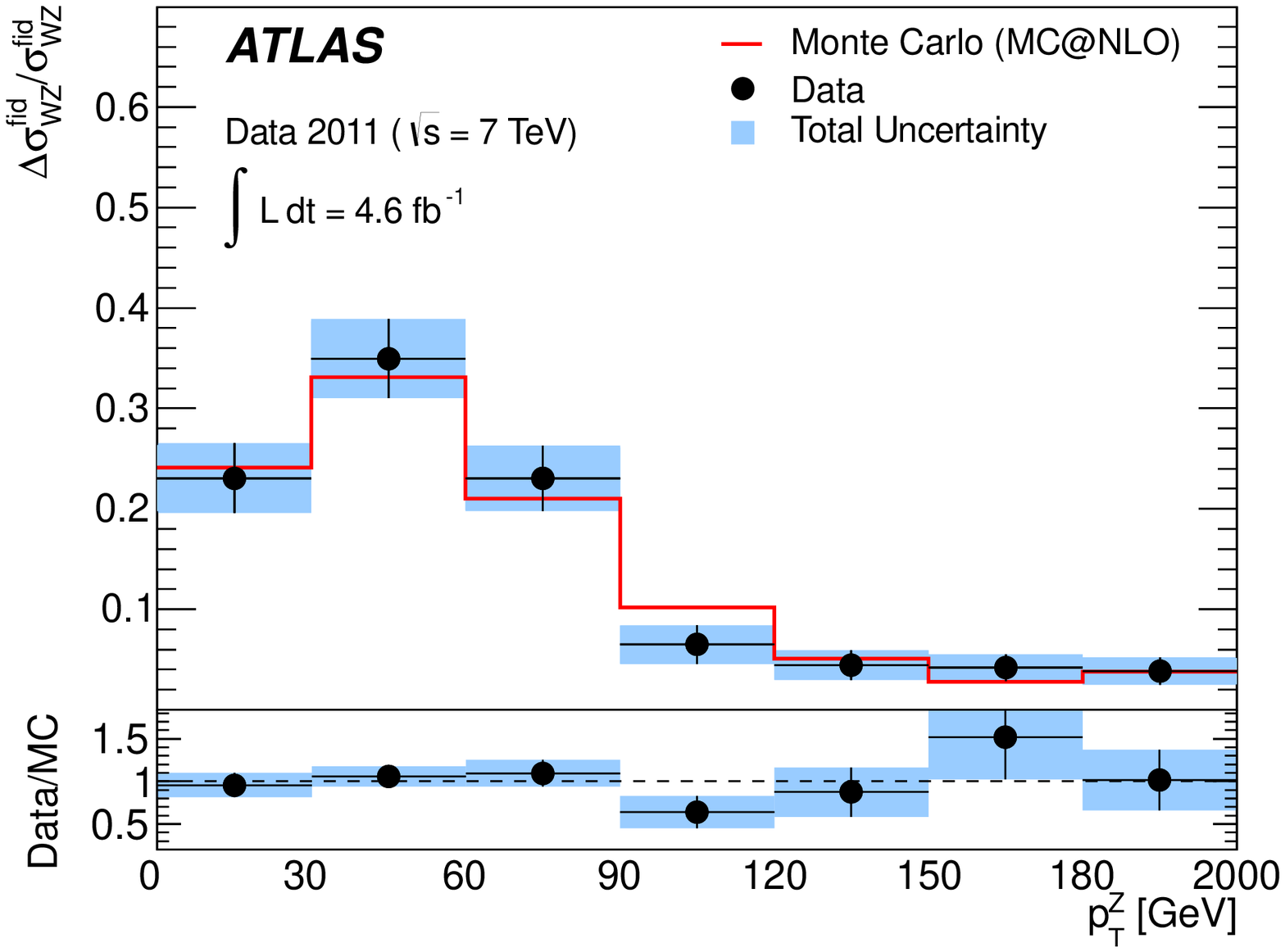}
    	\raisebox{0.6\columnwidth}{(b)}\includegraphics[width=0.9\columnwidth]{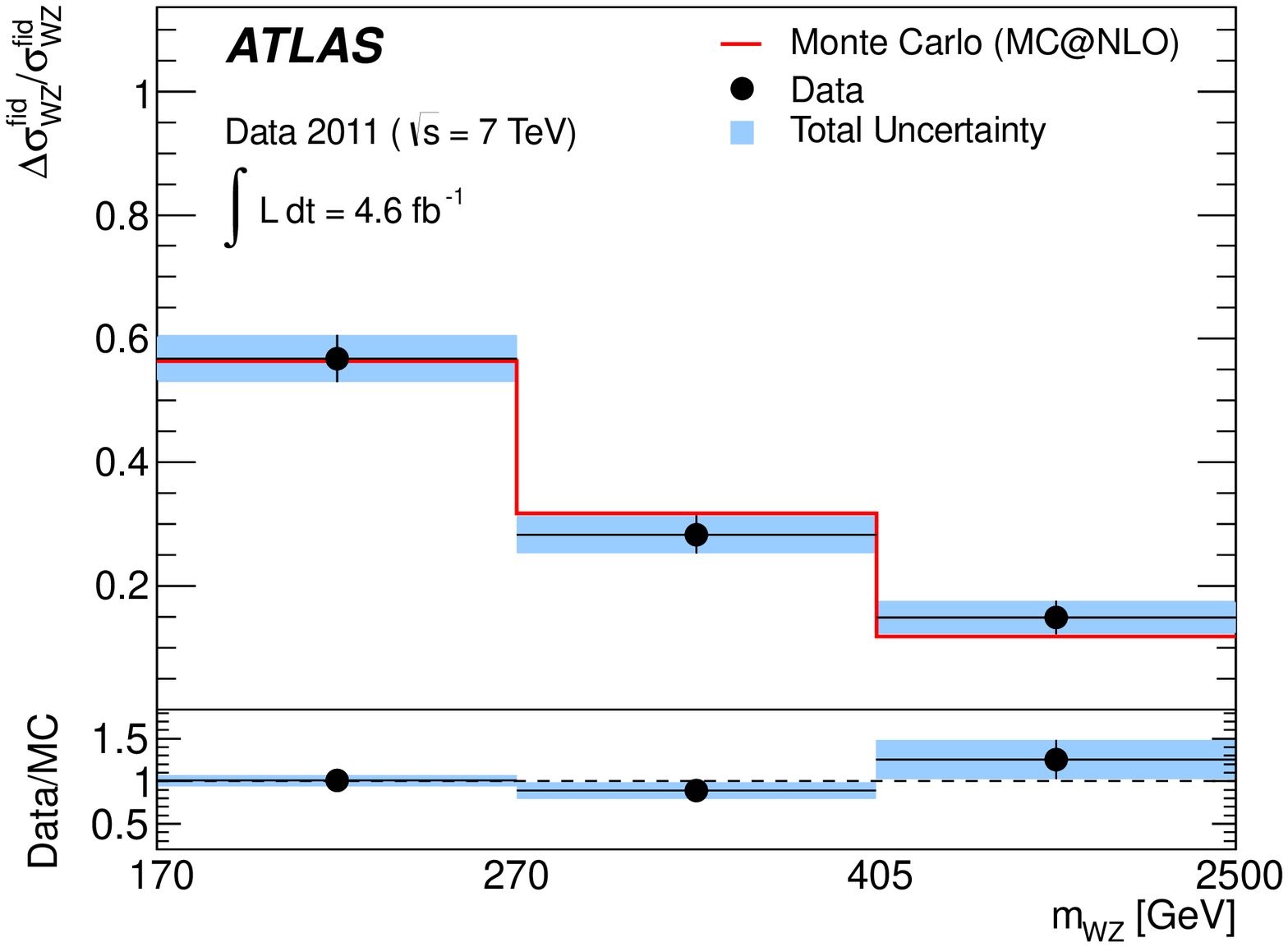}
        \caption{Normalized fiducial cross-sections $\Delta\xsecfid/\xsecfid$ in bins of (a) \ptZ\ and (b) \mWZ\
        compared with the SM prediction. The total uncertainty
        contains statistical and systematic uncertainties added in quadrature.}
        \label{Fig:Unfold:Data}
    \end{center}
\end{figure}

The dominant source of uncertainties on the normalized cross-sections is statistical.
The statistical uncertainties are determined by a Monte Carlo method.
Two thousand pseudo-experimental spectra are generated by fluctuating the content of each bin according to 
a Poisson distribution.
The unfolding procedure is applied to each pseudo-experiment, and the r.m.s.\ of the results
is taken as the statistical uncertainty.
The systematic uncertainties are evaluated by varying the response matrix for each source of the
uncertainty, and combining the resulting changes in the unfolded spectrum.
Because of the normalization, the results are affected only by the uncertainties that
depend on \ptZ\ or \mWZ.
The stability of the unfolding procedure is tested in two ways:
firstly by comparing the unfolded spectra after two and after three iterations,
and secondly by checking that the true variable distribution is correctly reproduced from a simulated
sample generated with non-zero anomalous couplings.

\begin{table}\centering
\caption{Normalized fiducial cross-sections and uncertainties in bins of \mWZ.}
\label{Tag:Unfold:Results1mWZ} 
\begin{tabular}{l>{$}c<{$}>{$}c<{$}>{$}c<{$}}
\hline
 \mWZ\ [GeV]                & [170,270]        & [270,405]        & [405,2500]      \\
\hline
 $\Delta\xsecfid(\mWZ)/\xsecfid$ & 0.568             & 0.283             & 0.149  \\
 Uncertainty & 0.038 & 0.030 & 0.027 \\
\hline
\end{tabular}
\end{table}

\section{Conclusion}\label{sec:Conclusion}
Measurements of $\WZ$ production in proton-proton collisions at
$\sqrt{s}=7$~TeV have been presented using 
a data sample with an integrated luminosity of \lumiRound\,\ifb, collected
with the ATLAS detector at the LHC.
The candidate $\WZ$ events were selected in the fully leptonic final states with
electrons, muons, and large missing transverse momentum.
In total, \WZNSigEventsObserved\ candidates were observed with a background
expectation of $\WZNBkgEventsRound\pm\WZNBkgEventsRoundErr$ events.
The fiducial and total cross-sections are determined to be
\[
\xsecfid = \WZfidXsec\WZfidXsecStatErr\text{(stat.)}\WZfidXsecSysErr\text{(syst.)}\WZfidXsecLumiErr\text{(lumi.)}\,\text{fb}, 
\]
and
\[
\xsectot = \WZtotXsec\WZtotXsecStatErr\text{(stat.)}\WZtotXsecSysErr\text{(syst.)}\WZtotXsecLumiErr\text{(lumi.)}\,\text{pb},
\]
respectively,
where the fiducial cross-section is defined by
$\pt^{\mu,e} > 15$\,\GeV\ for the leptons from the decay of the \Z\ bosons,
$\pt^{\mu,e} > 20$\,\GeV\ for the leptons from the decay of the \W\ bosons,
$|\eta^{\mu,e}| < 2.5$, 
$\pt^{\nu} > 25$\,\GeV, 
$|m_{\ell\ell}-m_{\rm Z}| < 10$\,\GeV,
$M_{\mathrm{T}}^{W} > 20$\,\GeV, and
$\Delta R > 0.3$ between the two leptons of all possible pairings of the three leptons.
These results are significantly more precise than the earlier ATLAS measurement~\cite{Aad:2011cx} which this paper supersedes.
The total cross-section is consistent with the SM expectation of $\WZtotXsecExpected\WZtotXsecExpectedErr$\,pb.
Limits on anomalous triple gauge couplings have been derived based on the
observed $\ptZ$ distribution. The 95\% confidence intervals are
\begin{eqnarray*}
\dgZ &\in& [\obsglow, \obsgup] \\
\dkZ &\in& [\obskappalow, \obskappaup] \\
\lamZ &\in& [\obslambdalow, \obslambdaup]
\end{eqnarray*}
without a form factor.
The limits are again more stringent than the earlier ATLAS measurement.
Normalized fiducial cross-sections have also been presented in bins of \ptZ\ and \mWZ,
 and are in good agreement with SM predictions.


\section*{Acknowledgements}

We thank CERN for the very successful operation of the LHC, as well as the
support staff from our institutions without whom ATLAS could not be
operated efficiently.

We acknowledge the support of ANPCyT, Argentina; YerPhI, Armenia; ARC,
Australia; BMWF, Austria; ANAS, Azerbaijan; SSTC, Belarus; CNPq and FAPESP,
Brazil; NSERC, NRC and CFI, Canada; CERN; CONICYT, Chile; CAS, MOST and NSFC,
China; COLCIENCIAS, Colombia; MSMT CR, MPO CR and VSC CR, Czech Republic;
DNRF, DNSRC and Lundbeck Foundation, Denmark; EPLANET and ERC, European Union;
IN2P3-CNRS, CEA-DSM/IRFU, France; GNAS, Georgia; BMBF, DFG, HGF, MPG and AvH
Foundation, Germany; GSRT, Greece; ISF, MINERVA, GIF, DIP and Benoziyo Center,
Israel; INFN, Italy; MEXT and JSPS, Japan; CNRST, Morocco; FOM and NWO,
Netherlands; RCN, Norway; MNiSW, Poland; GRICES and FCT, Portugal; MERYS
(MECTS), Romania; MES of Russia and ROSATOM, Russian Federation; JINR; MSTD,
Serbia; MSSR, Slovakia; ARRS and MVZT, Slovenia; DST/NRF, South Africa;
MICINN, Spain; SRC and Wallenberg Foundation, Sweden; SER, SNSF and Cantons of
Bern and Geneva, Switzerland; NSC, Taiwan; TAEK, Turkey; STFC, the Royal
Society and Leverhulme Trust, United Kingdom; DOE and NSF, United States of
America.

The crucial computing support from all WLCG partners is acknowledged
gratefully, in particular from CERN and the ATLAS Tier-1 facilities at
TRIUMF (Canada), NDGF (Denmark, Norway, Sweden), CC-IN2P3 (France),
KIT/GridKA (Germany), INFN-CNAF (Italy), NL-T1 (Netherlands), PIC (Spain),
ASGC (Taiwan), RAL (UK) and BNL (USA) and in the Tier-2 facilities
worldwide.

\bibliographystyle{spphys}
\bibliography{WZPaper}

\clearpage
\onecolumn
\begin{flushleft}
{\Large The ATLAS Collaboration}

\bigskip

G.~Aad$^{\rm 48}$,
T.~Abajyan$^{\rm 21}$,
B.~Abbott$^{\rm 111}$,
J.~Abdallah$^{\rm 12}$,
S.~Abdel~Khalek$^{\rm 115}$,
A.A.~Abdelalim$^{\rm 49}$,
O.~Abdinov$^{\rm 11}$,
R.~Aben$^{\rm 105}$,
B.~Abi$^{\rm 112}$,
M.~Abolins$^{\rm 88}$
% ...
\bigskip

% ...
$^{11}$ Institute of Physics, Azerbaijan Academy of Sciences, Baku, Azerbaijan\\
$^{12}$ Institut de F\'{i}sica d'Altes Energies and Departament de F\'{i}sica de la Universitat Aut\`{o}noma de Barcelona and ICREA, Barcelona, Spain\\
% ...
$^{21}$ Physikalisches Institut, University of Bonn, Bonn, Germany\\
% ...
$^{48}$ Fakult\"{a}t f\"{u}r Mathematik und Physik, Albert-Ludwigs-Universit\"{a}t, Freiburg, Germany\\
$^{49}$ Section de Physique, Universit\'{e} de Gen\`{e}ve, Geneva, Switzerland\\
% ...
$^{88}$ Department of Physics and Astronomy, Michigan State University, East Lansing MI, United States of America\\
% ...
$^{105}$ Nikhef National Institute for Subatomic Physics and University of Amsterdam, Amsterdam, Netherlands\\
% ...
$^{111}$ Homer L. Dodge Department of Physics and Astronomy, University of Oklahoma, Norman OK, United States of America\\
$^{112}$ Department of Physics, Oklahoma State University, Stillwater OK, United States of America\\
% ...
$^{115}$ LAL, Universit\'{e} Paris-Sud and CNRS/IN2P3, Orsay, France\\
% ...
\end{flushleft}


\end{document}